\documentclass{aa}
\usepackage{amsmath} 
\usepackage{txfonts}
\usepackage{graphicx} 
\usepackage{url}
\usepackage{natbib} 
\usepackage{ulem}
\newcommand{\el}[2]{$\rm{}^{#2}\kern-0.6pt#1$}
\newcommand{\elm}[2]{\rm{}^{#2}\kern-0.8pt\rm{#1}}%for math mode

\usepackage{color}
\definecolor{orange}{rgb}{1,0.5,0}

\usepackage[modulo,switch,pagewise]{lineno}

\usepackage{multirow}
\usepackage{threeparttable}
\begin{document}

\title{Models of red giants in the CoRoT asteroseismology fields combining asteroseismic and spectroscopic constraints}

\author{N. Lagarde\inst{1}, A. Miglio\inst{1,2}, P. Eggenberger\inst{3}, T. Morel\inst{4}, J. Montalb\'an\inst{5}, B. Mosser\inst{6}, T. S. Rodrigues\inst{5,7,8},\\L. Girardi\inst{7,8}, M. Rainer\inst{9}, E. Poretti\inst{9}, C. Barban\inst{6}, S. Hekker\inst{2,10,11}, T. Kallinger\inst{13}, M. Valentini\inst{14}, F. Carrier\inst{12}, \\M. Hareter\inst{13}, L. Mantegazza\inst{9}, Y. Elsworth\inst{1,2}, E. Michel\inst{6}, and A. Baglin\inst{6}}

\offprints{N.Lagarde, \email{lagarde@bison.ph.bham.ac.uk}}

\institute{School of Physics and Astronomy, University of Birmingham, Edgbaston, Birmingham B15 2TT, UK, email: lagarde@bison.ph.bham.ac.uk 
              \and Stellar Astrophysics Centre (SAC), Department of Physics and Astronomy, Aarhus University, Ny Munkegade 120, 8000 Aarhus C, Denmark
              \and Geneva Observatory, University of Geneva, Chemin des Maillettes 51, 1290 Versoix, Switzerland 
              \and Institut dÕAstrophysique et de G\'eophysique, Universit de Li\`ege, All\'ee du 6 Ao\^ut, B\^at. B5c, 4000 Li\`ege, Belgium 
              \and Departement of Physics and Astronomy G. Galilei, University of Padova, Vicolo dell'Osservatorio 3, I-35122 Padova, Italy
              \and LESIA, Observatoire de Paris, PSL Research University, CNRS, Universit\'e Pierre et Marie Curie, Universit\'e Denis Diderot,  92195 Meudon cedex, France 
             \and Osservatorio Astronomico di Padova INAF, Vicolo dell'Osservatorio 5, I-35122 Padova, Italy
              \and Laborat\'orio Interinstitucional de e-Astronomia, LIneA, Rua Gal. Jose Cristino 77, Rio de Janeiro, RJ, 20921-400, Brazil
              \and INAF - Osservatorio Astronomico di Brera, via E. Bianchi 46, 23807 Merate (LC), Italy
              \and Astronomical Institute ``Anton Pannekoek", University of Amsterdam, Science Park 904, 1098 XH Amsterdam, The Netherlands
              \and Max-Planck-Institut f\"ur Sonnensystemforschung, Justus-von-Liebig-Weg 3, 37077 G\"ottingen, Germany
              \and Katholieke Universiteit Leuven, Departement Natuurkunde en Sterrenkunde, Instituut voor Sterrenkunde, Celestijnenlaan 200D, 3001 Leuven, Belgium
              \and Institute for Astrophysics, University of Vienna, T\"urkenschanzstrasse 17, 1180 Vienna, Austria
              \and Leibniz-Institut f\"ur Astrophysik Potsdam (AIP), An der Sternwarte 16, 14482 Potsdam, Germany
}

\date{Received / Accepted 03 May 2015}

\authorrunning{N. Lagarde et al.} \titlerunning{Models of red giants in the CoRoT asteroseismology fields}

\abstract
{The availability of asteroseismic constraints for a large sample of red giant stars from the CoRoT and \textit{Kepler} missions paves the way
for various statistical studies of the seismic properties of stellar populations. }
{We use the first detailed spectroscopic study of 19 CoRoT red-giant stars (Morel et al 2014) to compare theoretical stellar evolution models to observations of the open cluster NGC 6633 and field stars.}
{In order to explore the effects of rotation-induced mixing and thermohaline instability, we compare surface abundances of carbon isotopic ratio and lithium with stellar evolution predictions. These chemicals are sensitive to extra-mixing on the red-giant branch. } 
{ We estimate mass, radius, and distance for each star using the  seismic constraints. We note that the Hipparcos and seismic distances are different. However, the uncertainties are such that this may not be significant.  Although  the seismic distances for the cluster members are self consistent they are somewhat larger than the Hipparcos distance. This is an issue that should be considered elsewhere.
Models including thermohaline instability and rotation-induced mixing, together with the seismically determined masses can explain the chemical properties of red-giants targets. However, with this sample of stars we cannot perform stringent tests of the current stellar models.
Tighter constraints on the physics of the models would require the measurement of the core and surface rotation rates, and of the period spacing of gravity-dominated mixed modes. A larger number of stars with longer times series, as provided by  \textit{Kepler} or expected with Plato, would help for ensemble asteroseismology.}
{}
{}


\maketitle

\section{Introduction}

The classical theory of stellar evolution fails to explain abundances anomalies observed in stars ascending the red giant branch (RGB).  Indeed, a large number of observations provide compelling evidence of an extra mixing process occurring when the low-mass stars reach the so-called bump in the luminosity function on the RGB. At that point, spectroscopic studies show a drop in the surface carbon isotopic ratio, and the lithium and carbon abundances, while nitrogen abundance increases slightly \citep[e.g.,][]{GiBr91, Gratton00, Tautvaisiene00, Smiljanic09,Tautvaisiene13}. 

A significant effort has been devoted to improving our understanding of the physical processes occurring in low- and intermediate-mass stars. The internal dynamics of these stars is altered by the effects of rotation, through the transport of both angular momentum and chemical species through the action of meridional circulation and shear turbulence, combined possibly with other processes induced by internal gravity waves or magnetic fields \citep[e.g.,][]{Zahn92, MaZa98, Eggenberger05,TalCha98, TalCha05, Charbonnel13}. \\

Rotation-induced mixing implies a variation of the chemical properties of stars during the main sequence and at the beginning of the RGB, successfully explaining many abundances patterns observed at the surface of low- and intermediate-mass stars  \citep{Palacios03, ChTa08, Smiljanic10, ChaLag10}. Rotation has also been investigated by several authors as a possible source of mixing during the RGB to explain abundances anomalies observed at the surface of RGB stars \citep[e.g.,][]{SwMe79, Charbonnel95, DeTo00, Palacios06}. However, the total diffusion coefficient of rotation during the RGB is too low to reproduce variations of chemical abundances on the first ascent giant branch as required by spectroscopic observations \citep[e.g.,][]{Palacios06}.\\

 Thermohaline instability driven by $^{3}$He-burning has been proposed as a process which is able to modify the photospheric compositions of bright low-mass red-giant stars \citep{ChaZah07a, ChaLag10}. This double diffusive instability is induced by the mean molecular weight inversion created, in these stars, by the $^{3}$He($^{3}$He,2p)$^{4}$He reaction (included in the pp-chains) in the thin radiative layer between the convective envelope and the hydrogen burning shell \citep{Eggleton06, Eggleton08,Lattanzio15}. This mechanism has a crucial impact on surface chemical properties of RGB stars in agreement with spectroscopic observations \citep{ChaLag10, Angelou11, Angelou12}. It is also very significant for the chemical evolution of light elements in our Galaxy \citep{Lagarde11, Lagarde12b}. \\
 
In summary, and as discussed in \citet{ChaZah07a}, \citet{ChaLag10}, and \citet{Lagarde11}, the effects of both rotation-induced mixing and thermohaline instability explain most of the spectroscopic observations of low- and intermediate-mass stars at various metallicities and evolutionary phases.\\

Hydrodynamic simulations, in 2D and 3D have been used to improve the constraints on the efficiency of thermohaline instability in stellar interiors \citep{Denissenkov09, Denissenkov10, DenissenkovMerryfield10, RosenblumGaraudetal11, Traxleretal11,Brown13}. These simulations currently show that double diffusive instability is not efficient enough to significantly change surface abundances \citep{Wachlin14}. However, they are still far from the parameter space relevant to the stellar regime. Future hydrodynamical simulations representative of conditions met in the stellar interior and taking the coupling of thermohaline instability with other mixing processes into account, will shed light on this discrepancy. \\

Additionally and independent of spectroscopy, the core rotation rate of red-giant stars measured by asteroseismology \citep[e.g.,][]{Beck12, Deheuvels12, Deheuvels14}, shows a significant disagreement with models predictions. It is clear that the physics of red-giant models should be improved in the light of new constraints brought by asteroseismology \citep{Eggenberger12, Marques13}. \\

Asteroseismology paves the way to a better understanding of stellar interiors. It provides valuable and independent constraints on current stellar models as well as on the physics of different transport processes. Indeed, for the first time, we have the possibility to determine the evolutionary state of red-giants \citep[e.g.,][]{Montalban10, Bedding10, Mosser11,Mosser14}, to estimate their core rotation rate \citep[e.g.,][]{Beck12, Deheuvels12, Deheuvels14}, and to deduce the properties of the core He-burning phase \citep{Mosser11, Montalban13}. Asteroseismology allows us to test stellar evolution models with more stringent constraints for clusters as well as for single stars.  With the CoRoT \citep{Baglin06} and \textit{Kepler} \citep{Borucki10} space missions, a large number of asteroseismic observations have been obtained for different kind of stars. They offer a unique opportunity to obtain some fundamental properties by observation of mixed modes in red giants \citep[e.g.,][]{ChMi13}. \\

To exploit all the potential of asteroseismic data from CoRoT and \textit{Kepler} missions, it is crucial to combine them with spectroscopic constraints \citep[e.g.,][]{Thygesen12, Molenda14}. The first spectroscopic study of the red giants lying in the asteroseismic CoRoT fields is \citet[][hereafter M14]{Morel14}.  It includes observations of the chemical tracers of extra-mixing in these stars (Li and carbon isotopic ratio). Moreover, this study gives surface chemical properties and seismic properties of three members of NGC 6633, which represents an ideal laboratory to study red-giant stars at the same age. \\

In this paper, we combine asteroseismic and spectroscopic measurements with stellar evolution models to use them to improve our knowledge of stellar interiors. 
We propose in this paper two complementary approaches to test model predictions of chemical transport with spectroscopic observations, and couple that to seismic constraints on stellar properties. In Sect. \ref{obs}, we briefly present the observed targets and discuss their stellar properties (radius, mass, and distance). We describe the physical input of the stellar evolution models in Sect. \ref{input}, while Sect. \ref{theo} includes a description of the rotation and thermohaline instability effects on the chemical properties of red-giant stars. Theoretical predictions are compared to spectroscopic (Li and $^{12}$C/$^{13}$C data) and asteroseismic observations in Sect. \ref{compa}, before the conclusion in Sect. \ref{conclu} 

\section{Stellar parameters}
\label{obs}

\begin{figure*} 
	\centering
		\includegraphics[angle=0,width=11.5cm, clip=true,trim=0cm 9cm 1cm 5cm]{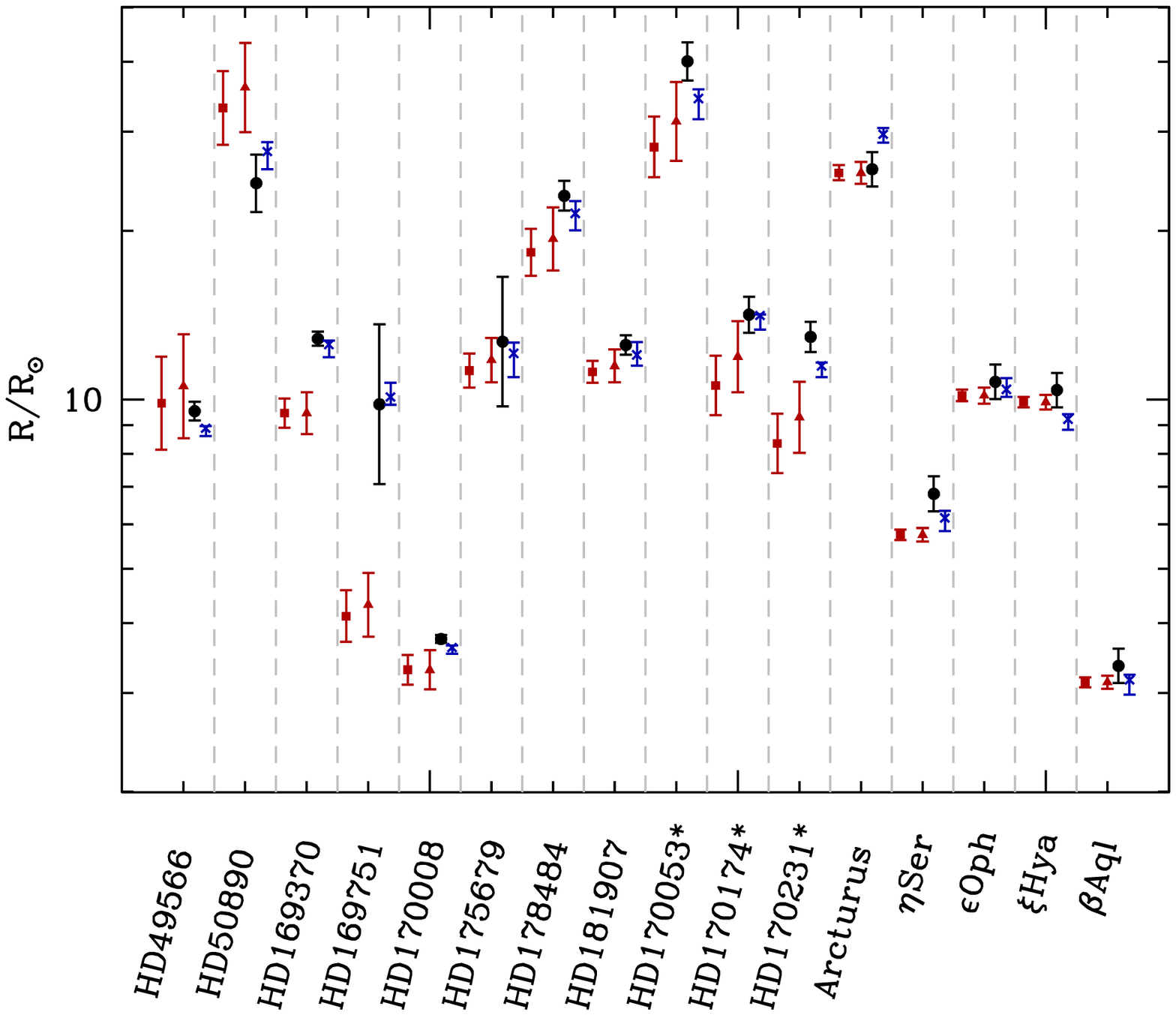}			 
		 \caption{Comparison between the radii obtained by different combinations of the available observational constraints. Radii are determined using asteroseismic constraints ($\nu_{\rm{max}}$,$\Delta \nu$) and T$_{\rm{eff}}$ (black dots), determined with PARAM (blue crosses), and using the parallax from the Hipparcos catalog \citep{vanLeeuwen07}, apparent magnitudes, and extinction from different prescriptions (red symbols). Squares represent radii computed with no extinction (A$_{\rm{V}}$=0), and triangles extinctions from the 3D Galactic extinction model by  \citet{Drimmel03}. Asterisks indicate stars that are members of NGC6633. }
	\label{sample_R}
\end{figure*}

\begin{figure*} 
	\centering
		\includegraphics[angle=0,width=11.5cm, clip=true,trim=0cm 9cm 1cm 5cm]{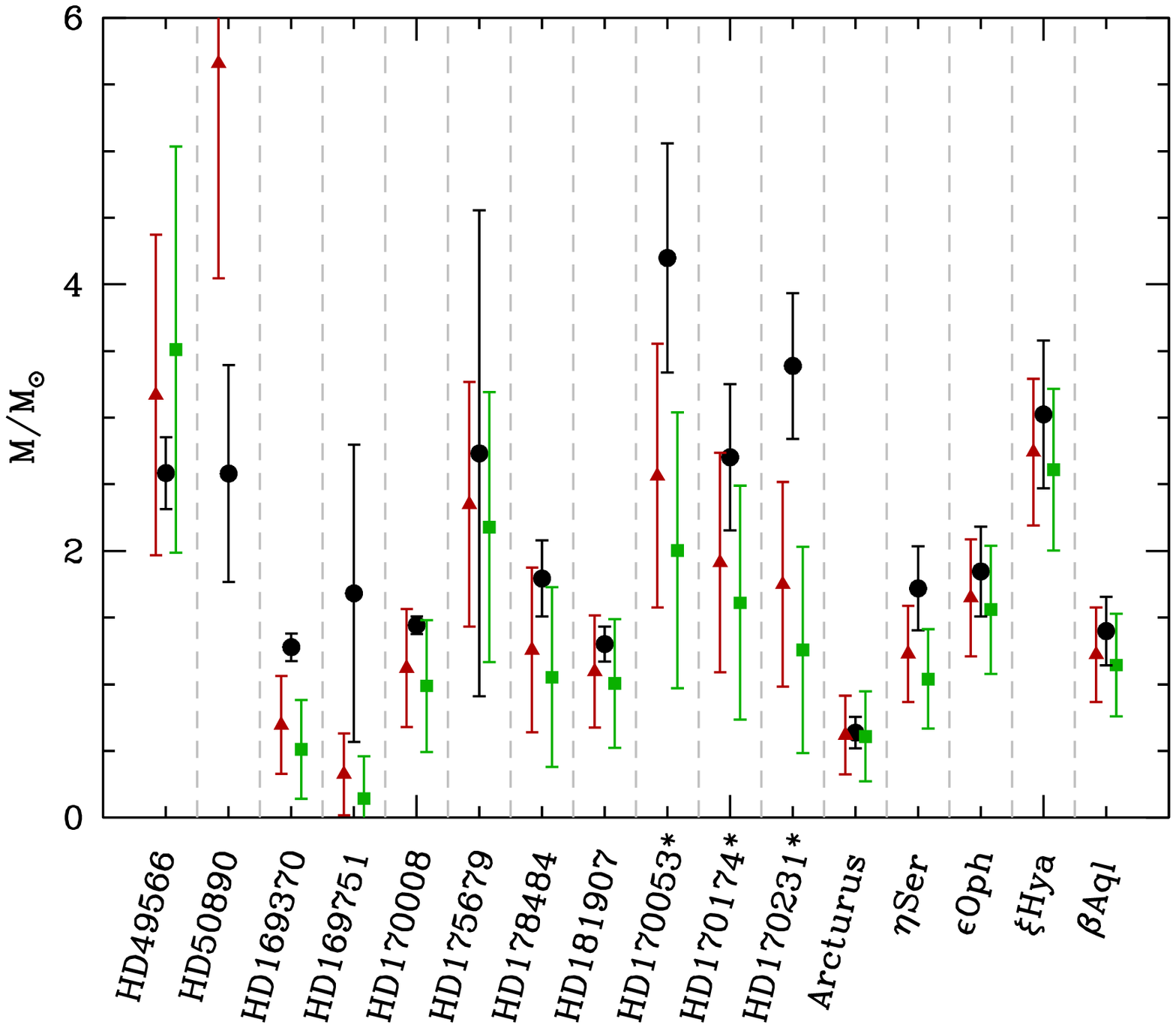}
			  \caption{Comparison between the masses obtained by different combinations of the observational constraints available. Masses are determined using asteroseismic constraints ($\nu_{\rm{max}}$,$\Delta \nu$) and T$_{\rm{eff}}$ (black dots), using $\nu_{\rm{max}}$, T$_{\rm{eff}}$, the extinction from Drimmel's model, and the stellar radius from the Hipparcos parallax (using T$_{\rm eff}$, red triangle), and using $\Delta \nu$ and the stellar radius from the Hipparcos parallax (using T$_{\rm eff}$) and stellar extinction by Drimmel's model (green square). Asterisks indicate stars that are members of NGC6633.}
	\label{sample_M}
\end{figure*}

\begin{figure*} 
	\centering
		\includegraphics[angle=0,width=11.5cm, clip=true,trim=0cm 6cm 1cm 5cm]{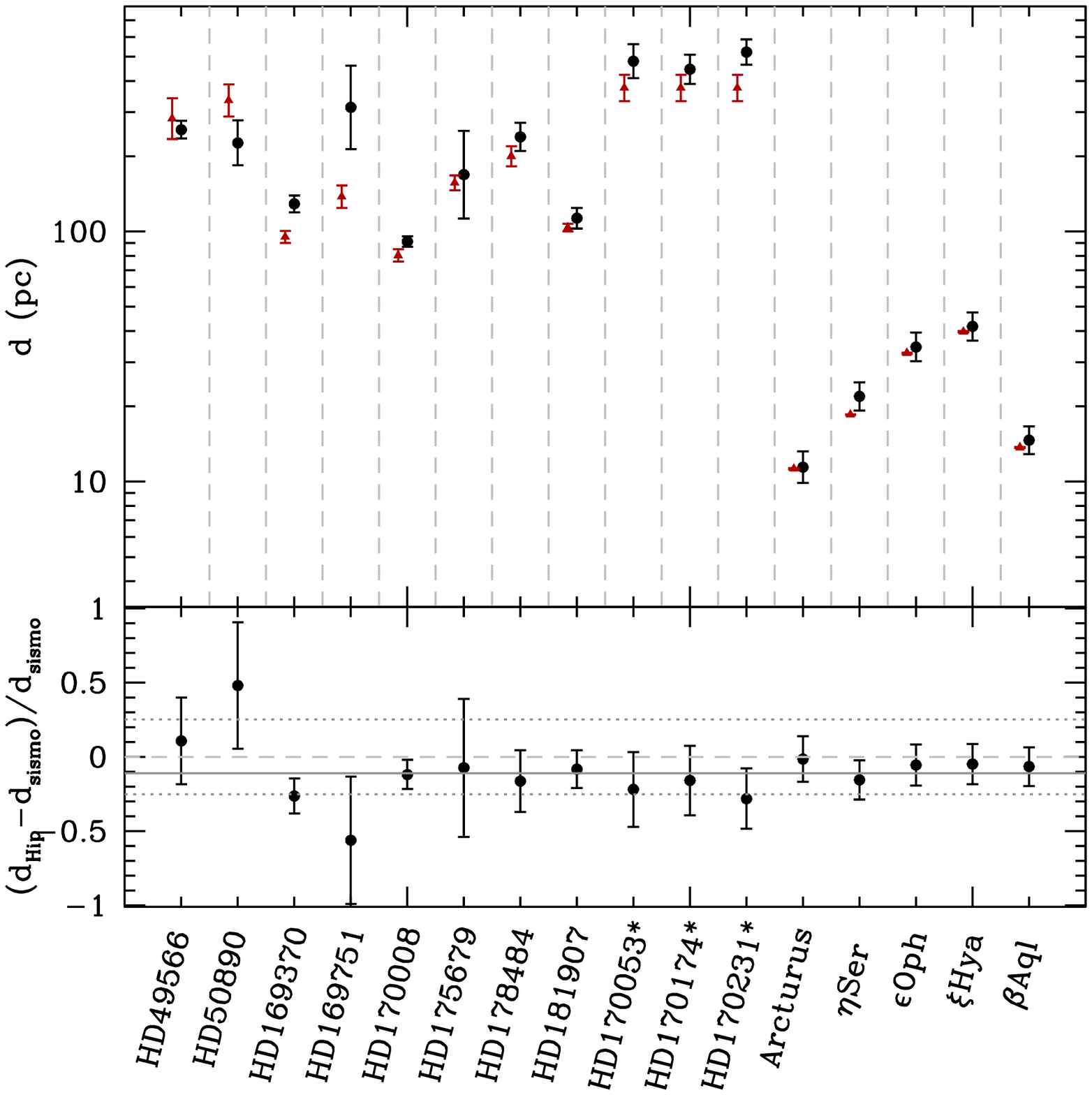}
			  \caption{Comparison between the distances determined from asteroseismic constraints ($\nu_{\rm{max}}$,$\Delta \nu$) and T$_{\rm{eff}}$ (black open circle), and the Hipparcos distance (red triangle). In the lower panel, the grey solid line represent the weighted average difference, while the grey dashed lines represent a difference of 25\%.  Asterisk identify the cluster members. }
	\label{sample_D}
\end{figure*}

In this section, we estimate the stellar radii and masses using several methods (spectroscopy, asteroseismology and astrometry) and then we discuss the differences between them. We use the spectroscopic determinations of chemical abundances published by M14. This sample is composed of 19 red-giant targets of which 15 were observed by CoRoT including three members of the young open cluster NGC 6633 (HD 170053, HD 170174, HD 170231, see tables \ref{dataseismo} and \ref{dataC}). M14 also derived the lithium abundances for all the stars in the sample and $^{12}$C/$^{13}$C for four of them (see table \ref{dataC}). The asteroseismic parameters of large separation, $\Delta\nu$, and frequency of maximum oscillation power, $\nu_{\rm{max}}$, are also taken from M14. Three different methods were used to obtain these global asteroseismic properties \citep[][]{MoAp09,Hekker10,Kallinger10}. See Sect. 4 of M14 for more details on the derivations of these parameters. The values are presented in table \ref{dataseismo}. We also use the effective temperature derived by M14 using the seismic-surface gravity as constraint. Astrometric parallaxes from Hipparcos \citep{vanLeeuwen07,vanLeeuwen09} have been used. \\

We start by determining radii and masses using seismic scaling relations (Eqs. \ref{M} and \ref{R}):

\begin{equation}
\frac{M}{M_{\odot}}\approx\left(\frac{\nu_{\rm{max}}}{\nu_{\rm{max,}\odot}}\right)^{3} \left(\frac{\Delta \nu}{\Delta \nu_{\odot}}\right)^{-4} \left(\frac{T_{\rm{eff}}}{T_{\rm{eff},\odot}}\right)^{3/2}
\label{M}
\end{equation}
\begin{equation}
\frac{R}{R_{\odot}}\approx\left(\frac{\nu_{\rm{max}}}{\nu_{\rm{max,}\odot}}\right) \left(\frac{\Delta \nu}{\Delta \nu_{\odot}}\right)^{-2} \left(\frac{T_{\rm{eff}}}{T_{\rm{eff},\odot}}\right)^{1/2}
\label{R}
\end{equation}
~\\ Solar reference values from M14 are $\Delta \nu_{\odot}$= 135.1 $\mu Hz$ , $\nu_{\rm{max,}\odot}$=3090 $\mu Hz$ and T$_{\rm{eff,\odot}}$=5777 K.\\

As the Hipparcos parallax, $\pi$, is available for most of the stars studied here, we can derive the stellar luminosity from the apparent magnitude in V-band, and then the stellar radius. The bolometric corrections are derived from \citet{Alonso99}. The extinctions are computed with the 3D Galactic extinction model by \citet{Drimmel03}. If the radius is known, the stellar mass can be estimated from only one of the two global seismic parameters $\Delta \nu$ (Eq.\ref{M2}) or $\nu_{\rm{max}}$ (Eq. \ref{M3}). This allows us to explore any systematic uncertainty on the mass determination.

\begin{equation}
\frac{M}{M_{\odot}}\approx \left(  \frac{\Delta \nu}{ \Delta \nu_{\odot}}\right)^2 \left(  \frac{R}{R_{\odot}}\right)^3 
\label{M2}
\end{equation}

\begin{equation}
\frac{M}{M_{\odot}}\approx \left(  \frac{ \nu_{\rm{max}}}{\nu_{\rm{max, \odot}}}\right) \left(  \frac{R}{R_{\odot}}\right)^2  \left(  \frac{T_{\rm{eff}}}{T_{\rm{ eff, \odot}}}\right)^{1/2}.
\label{M3}
\end{equation}

We can also derive the stellar radius from stellar models using both the spectroscopic and asteroseismic observational data. For this purpose, we use PARAM \citep{daSilva06,Rodrigues14} which computes the stellar properties with a Bayesian approach. This code compares observational data (T$_{\rm{eff}}$,[M/H],$\Delta \nu$,$\nu_{\rm{max}}$) with theoretical models (PARSEC isochrones, Bressan et al. 2012). 
It estimates also the distances and extinctions using observed magnitudes in several bandpass, bolometric corrections \citep[see references in][]{Rodrigues14}, and adopting a single extinction curve in terms of V-band. In our case, we use the UBVRI (from SIMBAD) and JHK \citep[2MASS,][]{2MASSref} bands, when available.

\begin{table}
  \caption{Seismic properties}
                      \label{dataseismo}  

	\scalebox{0.78}{
         \begin{threeparttable}
         \centering   

\begin{tabular}{|c | c | c | c | c | c | }
    \hline
      Name & T$_{\rm{eff}}$& $\Delta \nu$ & $\nu_{\rm{max}}$  & Mass$^{[1]}$& Radius$^{[1]}$ \\ 
   & (K) &($\mu Hz$) & ($\mu Hz$) & (M$_{\odot}$) & (R$_{\odot}$)  \\
     \hline
HD49566  & 5185 $\pm$ 50 & 7.37  $\pm$ 0.09 & 93 $\pm$ 2.78 & 2.6 $\pm$ 0.3 & 9.6 $\pm$  0.4  \\
HD50890   & 4710  $\pm$ 75 & 1.81 $\pm$ 0.065 & 15 $\pm$ 1.37 & 2.6 $\pm$  0.8 & 24.4 $\pm$  3  \\
HD169370  & 4520 $\pm$ 60   &  3.32 $\pm$ 0.03  &  27.2 $\pm$ 0.64 & 1.3 $\pm$  0.1 & 12.9 $\pm$  0.4  \\
HD169751  & 4910 $\pm$  55  & 5.7 $\pm$ 0.92  & 58.8  $\pm$ 1.63 & 1.7 $\pm$  1.1 & 9.8 $\pm$  3.2 \\
HD170008  & 5130 $\pm$ 50   & 22.4 $\pm$ 0.06 & 339 $\pm$ 4.58  & 1.5 $\pm$   0.1 & 3.7 $\pm$  0.1  \\
HD170031  & 4515 $\pm$ 65   & 3.87 $\pm$ 0.05 & 38.1 $\pm$ 0.77  & 1.9 $\pm$  0.15 & 13.3 $\pm$  0.4  \\
HD175679  & 5180 $\pm$ 50   & 4.94 $\pm$ 0.48 & 55.6 $\pm$ 9.74 & 2.8 $\pm$  1.8 & 12.7 $\pm$ 3.4  \\
HD178484  & 4440$\pm$ 60 & 1.63 $\pm$ 0.03 & 11.9 $\pm$ 0.52 & 1.8 $\pm$  0.3 & 23.2 $\pm$  1.4  \\
HD181907  & 4725 $\pm$ 65 & 3.48 $\pm$0.05 & 28.5 $\pm$ 0.74 & 1.3$\pm$  0.1 & 12.6 $\pm$  0.5  \\
HD170053 $^{[2]}$  & 4290 $\pm$ 65 & 1.09 $\pm$0.03 & 9.4$\pm$0.54 & 4.2 $\pm$  0.9  & 40.3 $\pm$  3.1  \\
HD170174 $^{[2]}$  & 5055 $\pm$ 55  & 4.16 $\pm$0.08 &44.6$\pm$2.7  & 2.7$\pm$ 0.6 & 14.2$\pm$ 1\\
HD170231$^{[2]}$   & 5175 $\pm$ 55 & 5.34$\pm$ 0.11 & 66.3 $\pm$ 2.96 & 3.4$\pm$ 0.6 &  13 $\pm$ 0.8 \\
$\alpha$ Boo & 4260  $\pm$ 60 & 0.82 $\pm$0.02 & 3.47 $\pm$ 0.17& 0.6 $\pm$  0.1 & 25.9$\pm$ 1.8  \\     
$\eta$ Ser   & 4935 $\pm$ 50  &  10 $\pm$ 0.25  & 125 $\pm$ 6.25 & 1.7$\pm$  0.3 & 6.8 $\pm$  0.5 \\
$\epsilon$ Oph  & 4940 $\pm$ 55  & 5.2 $\pm$ 0.13 & 53.5 $\pm$ 2.67 &  1.9 $\pm$  0.4 & 10.8  $\pm$  0.8  \\
$\xi$ Hya     & 5095 $\pm$ 50  &  7.0 $\pm$  0.175&  92.3 $\pm$ 4.61 &  3.1 $\pm$  0.5 & 10.4 $\pm$  0.7  \\
$\beta$ Aql  & 5110 $\pm$ 50  & 26 $\pm$ 0.65 &  410 $\pm$  20 & 1.4 $\pm$  0.3 & 3.4 $\pm$  0.2 \\
   \hline
 \end{tabular}
 
        \begin{tablenotes}
        \item[1] computed from the seismic scaling relations. 
        \item[2] NGC 6633 members
       	\end{tablenotes}
	
	\end{threeparttable}
	}

\end{table}

Having described the different methods used, we now present a comparison between radii and masses determined using the different combinations of seismic and non-seismic constraints. Radii and masses are computed using Eqs. \ref{M} and \ref{R}, and obtained by the other methods (Eqs. \ref{M2}, \ref{M3}, and parallax). We also consider extinctions, A$_{\rm{V}}$, using  \citet{Drimmel03} as described above, and A$_{\rm{V}}$=0. These results are presented in Figs. \ref{sample_R} and \ref{sample_M}.\\

The radii determined with PARAM have the best precision, due to the bayesian approach and the use of priors on stellar evolutionary tracks. In most cases, masses and radii determined using different methods agree within standard uncertainty. However, this is not a very stringent test since the typical uncertainties are $\sim$9\,\% for the radii and $\sim$22\,\% for the masses.

Fig. \ref{sample_M} shows a good agreement between red and green symbols. This is mostly due to the strong dependency of the mass estimates (from Eqs. \ref{M2} and \ref{M3}) on the stellar radius, which in both cases is derived using Hipparcos parallaxes.  \citet{Miglio12b} suggested that a relative correction to the scaling relation should be considered between red clump stars and RGB stars, affecting the mass determination of clump stars by $\sim$ 10\%. However, the uncertainties observed here are larger than this correction. Similarly, the correction proposed by \citet{Mosser13a} are smaller (6\% for the stellar radius and 3\% for the mass). \\
 
Figure \ref{sample_D} displays the distance to each star as given in the Hipparcos catalog, and computed via asteroseismic constraints. The lower panel shows the relative difference between the two quantities. The weighted average of the relative difference between seismic and Hipparcos distances is $-$0.12 with a statistical uncertainty of 0.03. We used the approach developed by \citet{Chaplin98} \citep[and used in][]{Miglio12b} to compute the weighted average of the relative distance and include the Student t-distribution to take into account the small number of points. This marginally significant differences suggest that the seismic distances are overestimated compared to the Hipparcos distances, with consequences for the seismic radii and masses (see Figs. \ref{sample_R} and \ref{sample_M}).

The weighted average of the relative difference between Hipparcos and seismic distances for the stars in the cluster is -0.23 $\pm$ 0.10. The accuracy of Hipparcos parallaxes has recently been questioned by \citet{Meilis14} in the case of the Pleiades, suggesting that the Hipparcos distance is overestimated by $\sim$12 \%. However, our current poor knowledge of systematic uncertainties on the seismically determined distances prevents us from contributing to this debate.

\section{Theoretical predictions }
\label{TH}
\subsection{Input physics of the hydrodynamical stellar models }
\label{input}

In this paper, we use the stellar evolution code STAREVOL. In a series of three papers \citep[][hereafter CL10, L11, and L12a respectively]{ChaLag10, Lagarde11, Lagarde12a}, the effects of rotation-induced mixing and thermohaline instability on the structure, evolution, nucleosynthesis as well as on global asteroseismic properties of low- and intermediate-mass stars at various metallicities, were discussed. A detailed description of the input physics is given in L12a. The treatment of convection is based on a classical mixing-length formalism with $\alpha_{MLT}=$1.6. \\ 

The treatment of rotation-induced mixing follows the complete formalism developed by \citet{Zahn92, MaZa98} (see for more details CL10 and L12a). Solid-body rotation is assumed when the star arrives on the zero age main sequence (ZAMS). Typical initial (i.e., ZAMS) rotation velocities are chosen depending on the stellar mass based on observed rotation distributions in young open clusters \citep{Gaige93}. In these models, we consider that the turbulent diffusivity related to thermohaline instability induced by $^{3}$He burning develops as long thin fingers with aspect ratio consistent with predictions by \citet{Ulrich72} and confirmed by the laboratory experiments \citep{Krish03}. We adopt an aspect ratio of 5, which nicely reproduces the observed chemical properties of red-giant stars \citep[][and CL10]{ChaZah07a}, and allows us to solve the so-called ``$^{3}$He problem'' in the Milky Way \citep{Lagarde12b}.\\

These new stellar models produced by STAREVOL, including thermohaline instability as well as rotation-induced mixing, can explain the behaviour of different chemical abundances in main-sequence, and red-giant stars observed in the field and open cluster stars over the relevant mass and metallicity range \citep[][CL10, and Lagarde et al. 2015 in prep.]{Smiljanic10}.\\

In the following sections, we briefly present the effects of rotation-induced mixing and thermohaline mixing on the surface abundances of lithium and carbon isotopic ratio drawing the stellar evolution. We also present a comparison, in Sect.4, between our predictions from L12a and the data for the CoRoT red giants targets.

\subsection{Evolution of Lithium and isotopic ratio of carbon.}
\label{theo}

\begin{figure} 
	\centering
		\includegraphics[angle=0,width=0.5\textwidth, clip=true,trim=1cm 6.5cm 1cm 4cm]{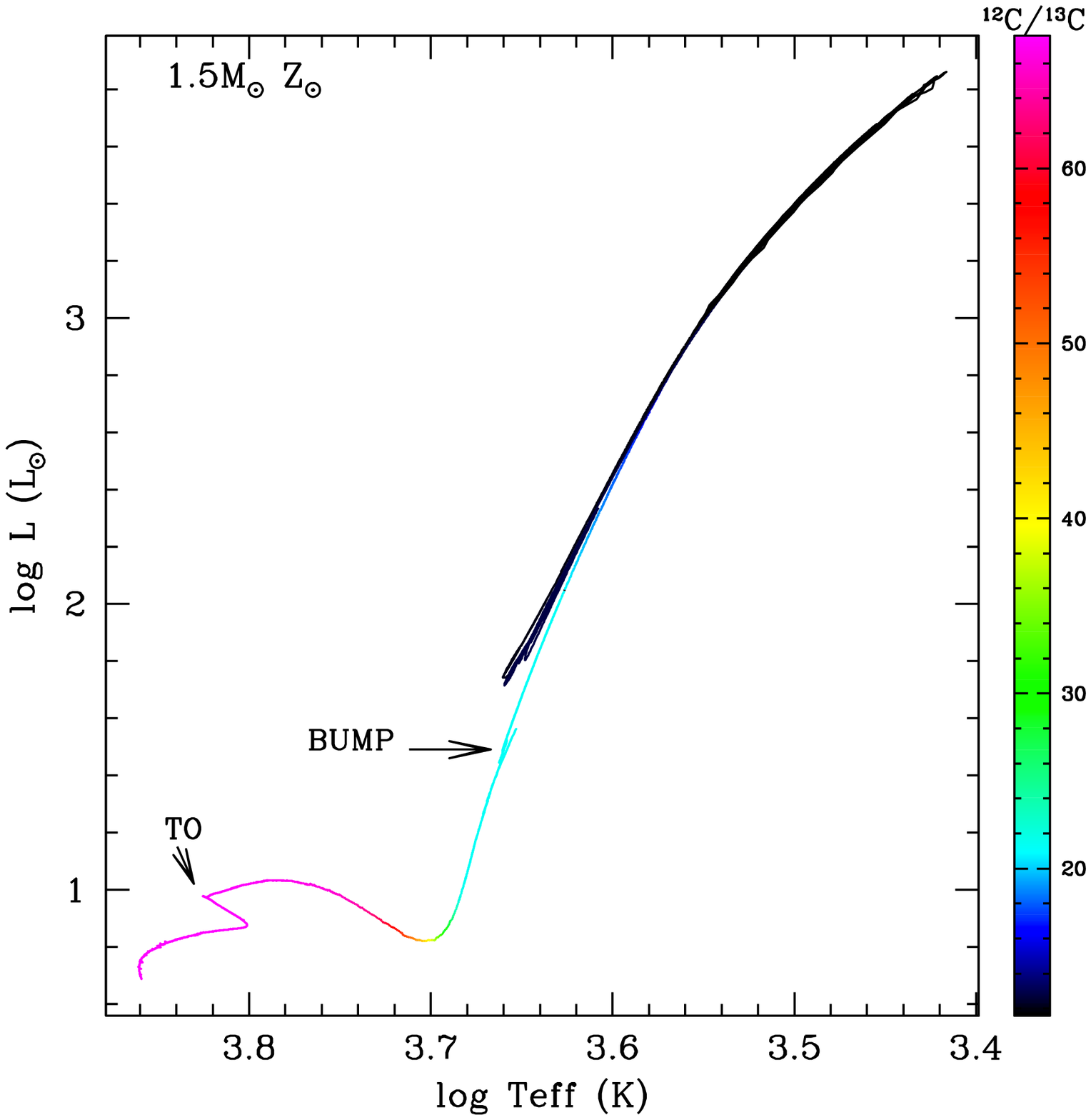}
			  \caption{HR diagram of model with $M$ = 1.5 M$_{\odot}$ and solar metallicity including the effects of rotation-induced mixing and thermohaline instability. The color represents the mass fraction of $^{12}$C/$^{13}$C at the surface. The turn-off and the bump luminosity are indicated on the figure.} 
	\label{HR}
\end{figure} 

\begin{figure} 
	\centering
		\includegraphics[angle=0,width=0.5\textwidth, clip=true,trim=2cm 6.5cm 1cm 4cm]{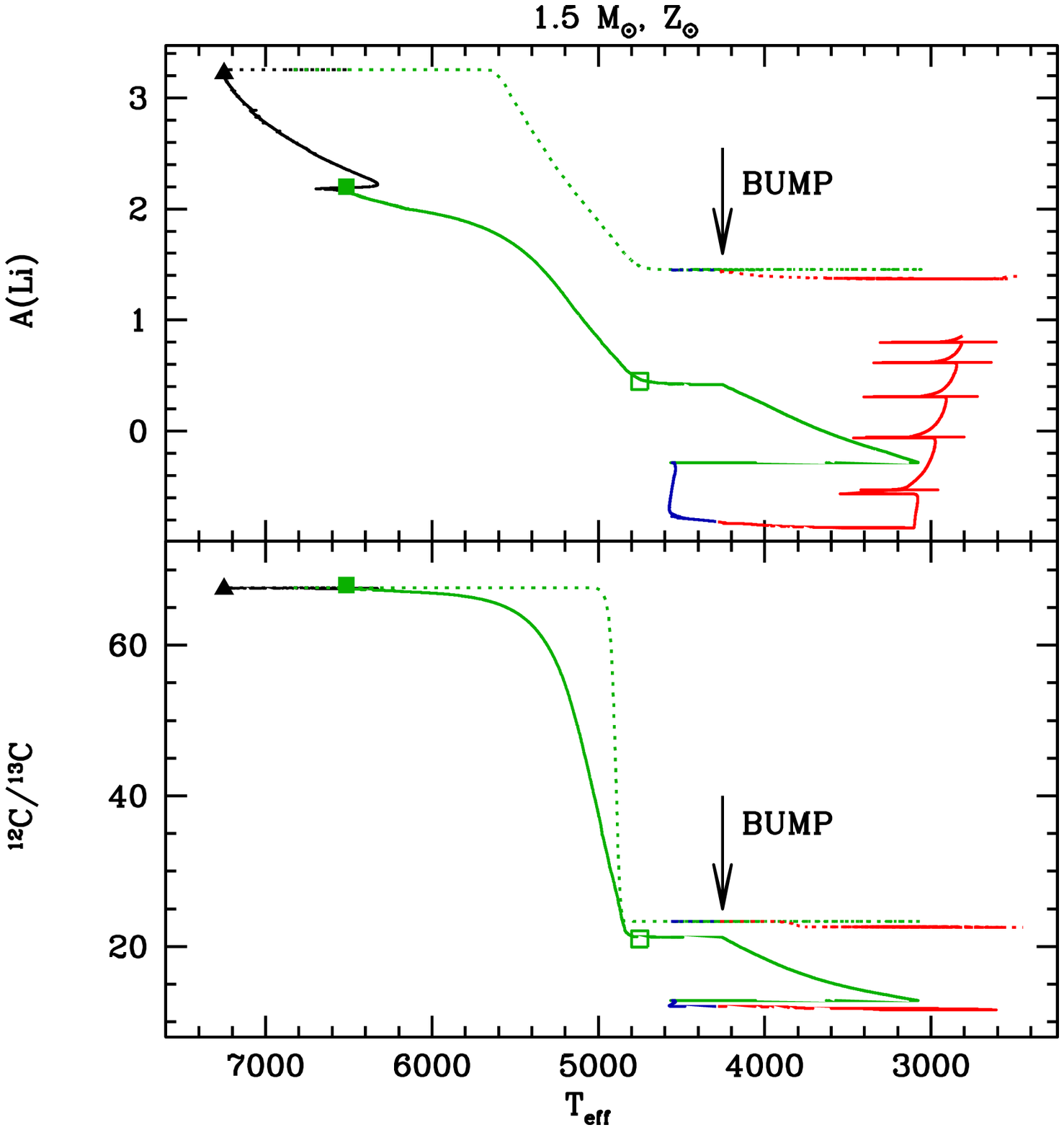}
			  \caption{Evolution of the lithium surface abundances (A(Li)=log($\frac{X(Li)}{X(H)}\frac{A_H}{A_{Li}})+12$), with X(Li) the lithium mass fraction), and carbon isotopic ratio as a function of effective temperature. Predictions are shown for 1.5 M$_{\odot}$ models at solar metallicity computed following standard prescriptions (dotted line) and including rotation-induced mixing as well as thermohaline instability (V$_{\rm{ZAMS}}$=110 km/s, (solid line) from the ZAMS up to the TP-AGB phase. Evolutionary phases are indicated by a colour label: main sequence (black), red-giant branch (green), He-burning phase (blue), and asymptotic giant branch (red). ZAMS (black triangle) and first dredge-up (start/end ; full/open squares) are indicated. } 
	\label{thepredic}
\end{figure}

\begin{figure} 
	\centering
		\includegraphics[angle=0,width=0.5\textwidth, clip=true,trim=1cm 6.5cm 1cm 4cm]{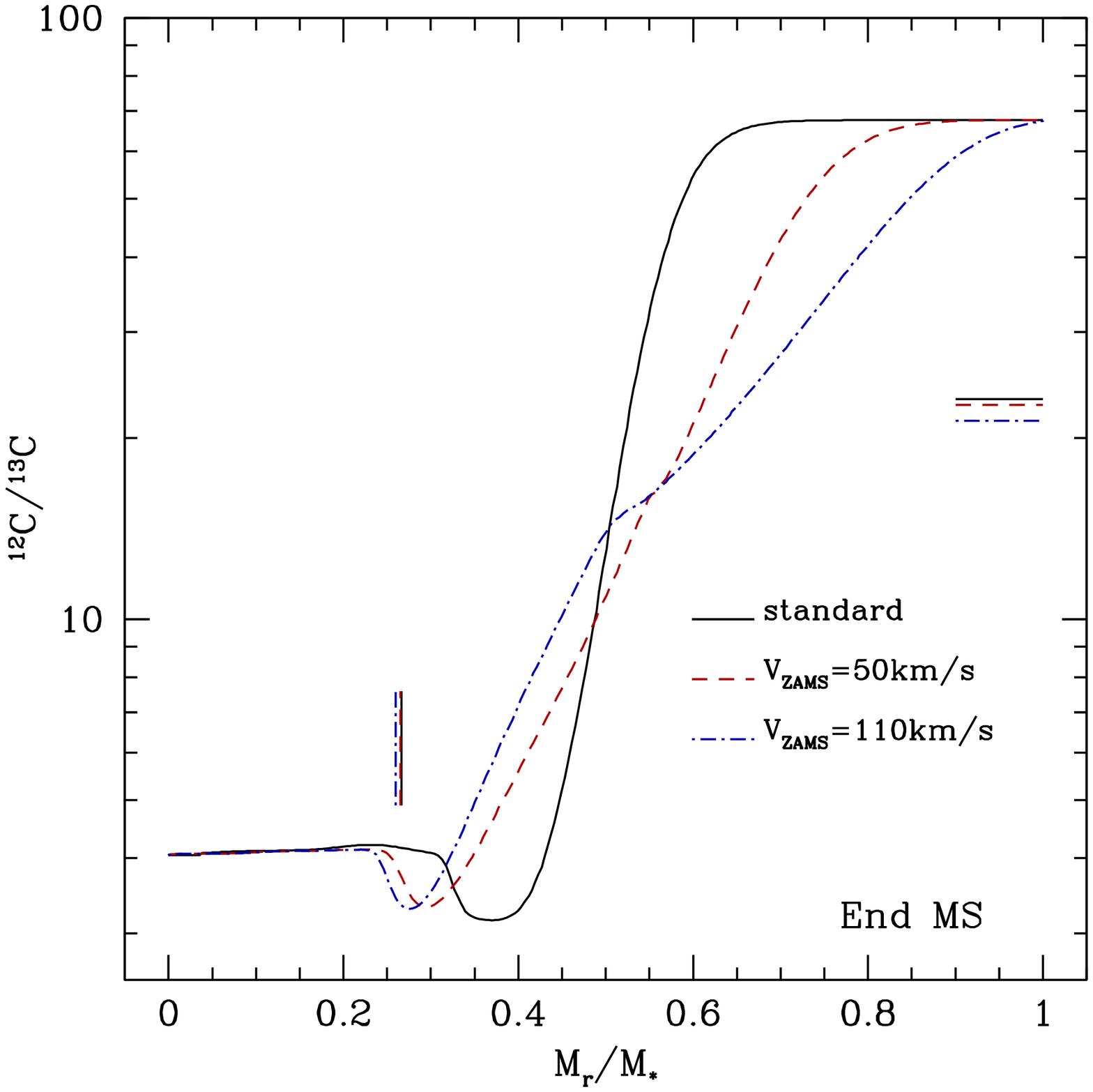}
			  \caption{Chemical profiles of isotopic ratio of carbon as a function of coordinate in mass (M$_{r}$/M$_{\odot}$) at the turn-off of the 1.5 M$_{\odot}$ star computed with different initial rotation velocities. The black solid, red dashed, and blue dot-dashed lines represent standard models, models including rotation with V$_{\rm{ZAMS}}$ = 50 km/s, and 110 km/s, respectively. The vertical lines show, in all cases, the maximum depth reached by the convective envelope at its maximum extent during the first dredge-up, while the horizontal lines indicate the surface values of $^{12}$C/$^{13}$C at the end of the first dredge-up.} 
	\label{profTO}
\end{figure} 

\begin{figure} 
	\centering
		\includegraphics[angle=0,width=0.5\textwidth, clip=true,trim=1cm 6.5cm 1cm 4cm]{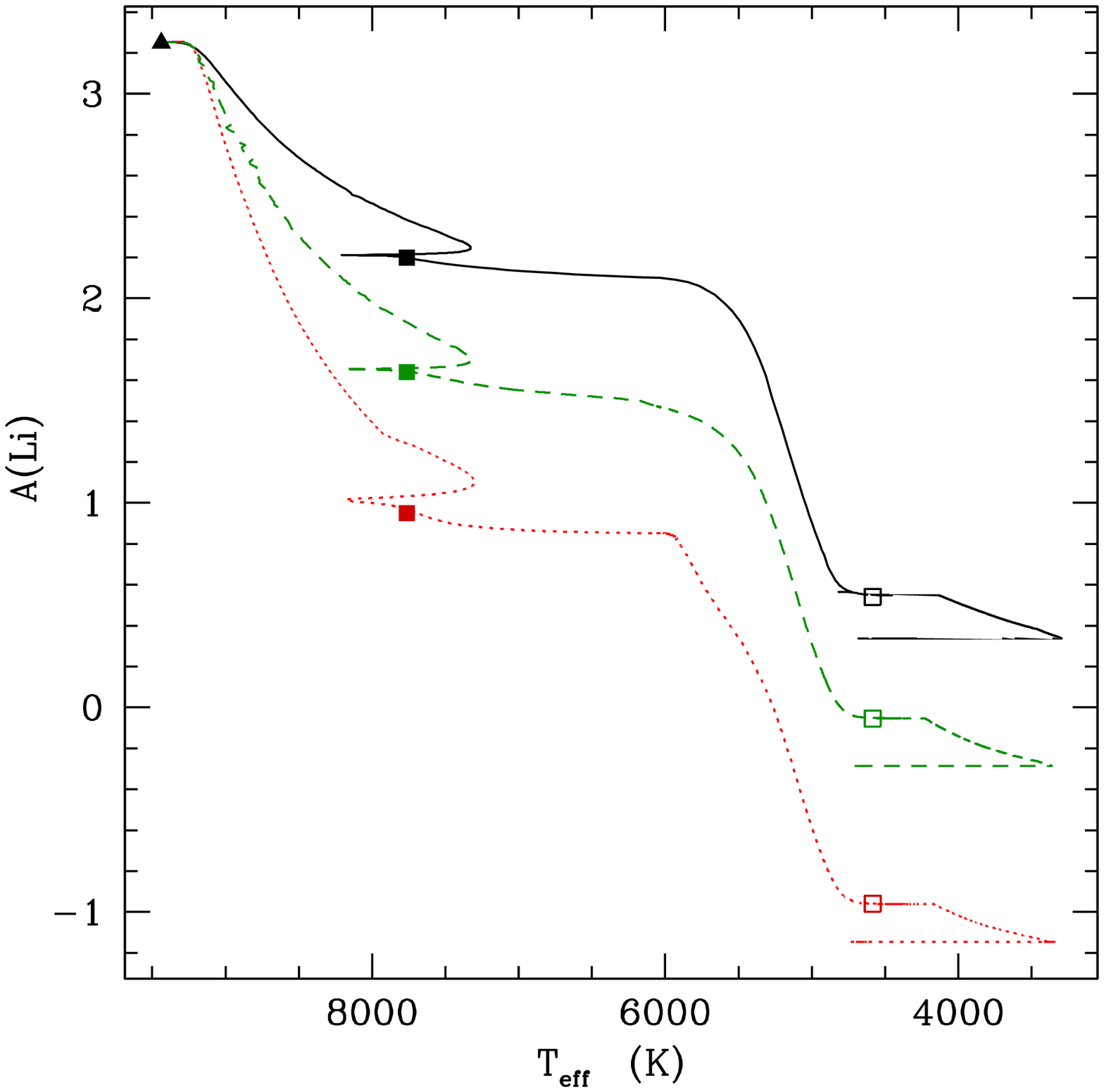}
			  \caption{Evolution of the surface abundances of Lithium, A(Li) as a function of effective temperature. Predictions are shown for 2.0 M$_{\odot}$ models at solar metallicity computed including rotation-induced mixing (V$_{\rm{ZAMS}}$=110, 180, 250 km/s, black solid, green dashed, and red dotted lines respectively) from the ZAMS up to the core-He-burning phase. ZAMS (black triangle) and first dredge-up (start/end ; full/open squares) are indicated.  }
	\label{thLi2p0}
\end{figure} 

\begin{figure} 
	\centering
		\includegraphics[angle=0,width=0.5\textwidth, clip=true,trim=1cm 6.5cm 1cm 4cm]{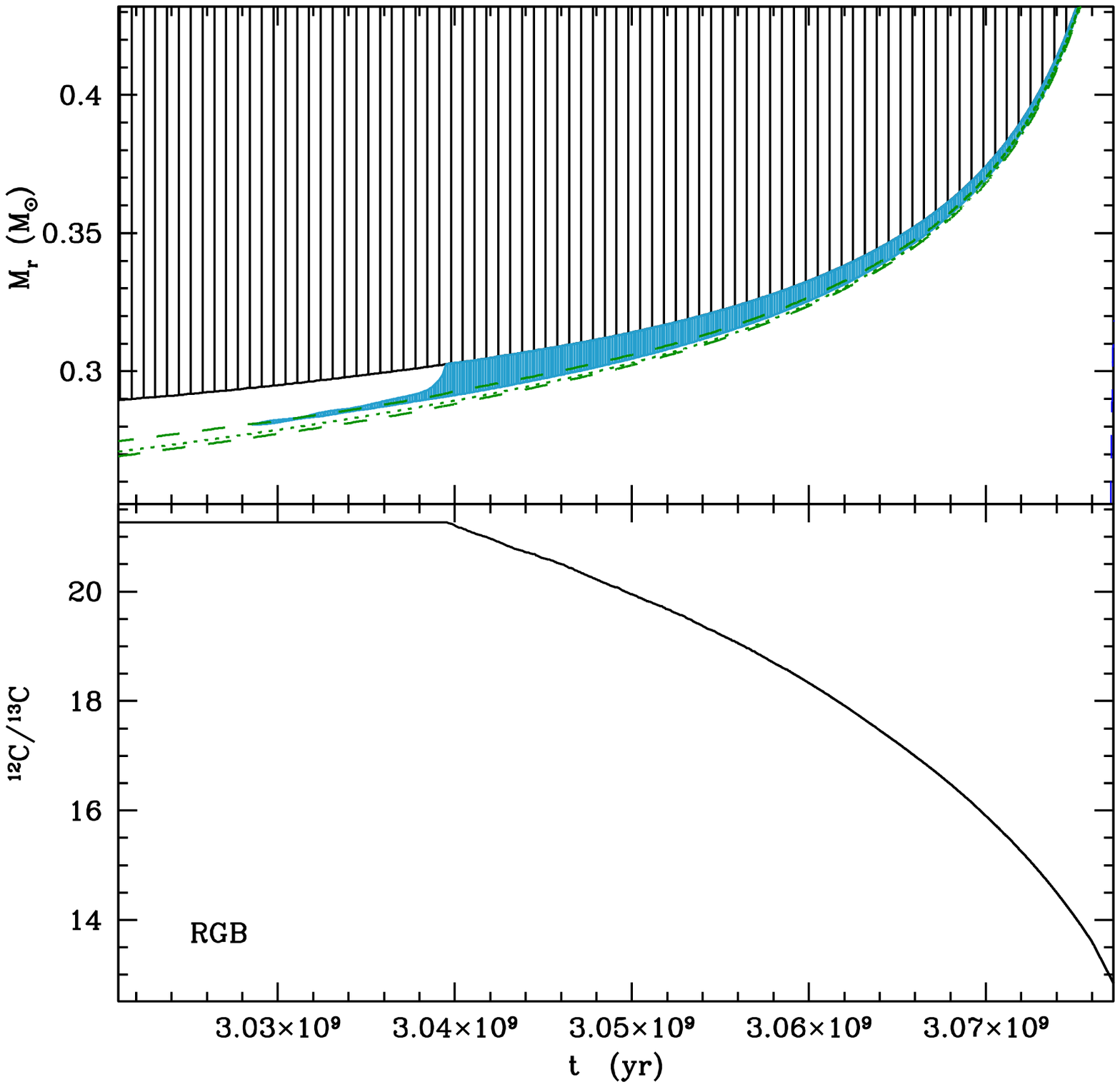}
			  \caption{\textit{Top panel:} Kippenhahn diagram for the 1.5 M$_{\odot}$ star computed with thermohaline instability and rotation-induced mixing. Here we focus on the red-giant branch. Green dashed lines delimit the hydrogen-burning shell above the degenerate helium core, and the
dotted line shows the region of maximum nuclear energy production. The region where thermohaline instability takes place is indicated in blue. \textit{Bottom panel: }The evolution of $^{12}$C/$^{13}$C during the RGB.} 
	\label{kippenRGB}
\end{figure}

We focus on the effects of rotation-induced mixing and thermohaline instability on Li and $^{12}$C/$^{13}$C surface abundances. The best indicator of non-standard transport processes in evolved low-mass stars is the carbon isotopic ratio, as discussed in literature and supported by numerous observations \citep[e.g.,][]{Gilroy89,Gratton00,Smiljanic09}. The most fragile element, lithium, burns by proton captures at relatively low temperature ($\sim$ 2.5.10$^6$ K) and is preserved only in the most external stellar layers \citep[e.g.,][]{Pasquini04, ChaTal99, Palacios03}. $^{12}$C/$^{13}$C and Li can be useful in the characterization of mixing mechanisms because they burn at different temperatures, i.e. at different depths in stellar interiors.

Figure \ref{HR} presents the evolutionary track in the Hertzsprung-Russell diagram (HRD) of the 1.5 M$_{\odot}$ model computed with thermohaline instability and rotation (V$_{\rm{ZAMS}}$=110 km/s). We select two points along the track (turn-off and the bump luminosity) to discuss the evolution of carbon isotopic ratio and lithium at the surface of low- and intermediate-mass stars. 

Figure \ref{thepredic} represents the evolution of lithium and isotopic ratio of carbon as a function of effective temperature for the same model. Standard predictions (i.e. non-rotating models without thermohaline mixing) are also shown. \\

In the standard case (solid line on Fig. \ref{thepredic}), the surface depletion of both Li and $^{12}$C/$^{13}$C begins at relatively low T$_{\rm{eff}}$ (i.e. at T$_{\rm{eff}}\sim$ 5600 K for Li and T$_{\rm{eff}}\sim$5000 K for $^{12}$C/$^{13}$C). This is due to the dilution of external layers when the convective stellar envelope deepens in mass during the first dredge-up. The surface abundances are not predicted to change after the end of the first dredge-up until the star reaches the early-AGB. We will discuss next the effects of rotation-induced mixing and thermohaline instability on the surface abundances of isotopic ratio of carbon and lithium. \\

\subsubsection{At the turnoff}

The modifications of the stellar internal and surface chemical abundances are driven by rotation-induced mixing on the main sequence and convective dilution during the first dredge-up episode on the sub-giant branch and early-RGB.
Figure \ref{profTO} describes the effect of different initial velocities on the $^{12}$C/$^{13}$C profile as a function of coordinate in mass\footnote{M$_{r}$/M$_{\odot}$ allows to point out the central regions of the star.} (M$_{r}$/M$_{\odot}$), at the end of central hydrogen burning in a 1.5 M$_{\odot}$ model. In the rotating models, the abundance gradients are smoothed compared to the standard case. Rotation-induced mixing modifies the internal chemical structure of main-sequence stars, enlarging the Li-free regions and modifying the $^{13}$C and $^{12}$C profiles. \\

\subsubsection{Toward the red-giant branch}

When the star expands after the turnoff (TO) and evolves toward the red-giant branch, its convective envelope deepens and engulfes most of the regions that have been nuclearly processed as indicated by the vertical lines on Fig. \ref{profTO}. This leads to the first dredge-up which changes the surface chemical properties of the star. Rotating sub-giant stars have lower surface carbon isotopic ratio and lithium abundance compared to standard models \citep[see Fig.\ref{thepredic}, e.g.,][]{Palacios03, Smiljanic09}. Indeed, when the initial velocity increases, the surface abundance of lithium after the first dredge-up decreases (see Fig. \ref{thLi2p0}, at T$_{\rm{eff}}\sim$ 4800 K). This implies that two stars with the same global properties (luminosity, surface gravity, or effective temperature) could have a very different surface abundance of lithium due to their very different rotation velocities.\\

\subsubsection{red-giant branch}

The $^{12}$C/$^{13}$C and Li abundances at the surface decrease during the RGB, and specifically at the bump luminosity, as shown by models including rotation-induced mixing and thermohaline instability (Fig.\ref{HR} and Fig.\ref{thepredic}). 
Indeed, as already discussed by \citet{ChaZah07a} and CL10, on the RGB, thermohaline mixing becomes efficient close to the bump luminosity (T$_{\rm{eff}}\sim$ 4200 K). Then, the theoretical abundances of Li and $^{12}$C/$^{13}$C drop again (for more details see CL10), while they would stay constant in the standard case. \\
Figure \ref{kippenRGB} displays a Kippenhahn diagram for a 1.5M$_{\odot}$ model,  as well as the evolution of the surface abundance of carbon isotopic ratio, during the RGB.  As discussed in \citet{ChaZah07a} and CL10, thermohaline instability develops at the top of the hydrogen burning shell (HBS) by an inversion of mean molecular weight. This instability is induced by the $^{3}$He($^{3}$He,2p)$^{4}$He reaction which takes place in the HBS, only after the star has reached the luminosity bump. This occurs when the HBS crosses the molecular weight discontinuity left behind by the first dredge-up (at t$\sim$ 3.028.10$^{9}$ yrs on Fig. \ref{kippenRGB}). As soon as the thermohaline instability sets in, newly emitted  protons diffuse outward, spreading out the molecular-weight inversion and enlarging the thermohaline region until it reaches the convective envelope (at t$\sim$ 3.039 10$^{9}$ yr on Fig. \ref{kippenRGB}). As a consequence, $^{12}$C and $^{13}$C diffuse respectively inwards and outward, leading to a decrease of the surface carbon isotopic ratio. The surface Li abundance decreases too (Fig. \ref{thepredic}). In addition, rotation-induced mixing leads to an earlier (in terms of luminosity) beginning of thermohaline instability on RGB compared to a model without rotation (see CL10). \\

\subsubsection{He-burning phase}

After the star reaches the RGB tip (at T$_{\rm{eff}}\sim$ 3100 K), its effective temperature increases until it settles in the red clump (blue part on Fig. \ref{thepredic}) before decreasing again when it starts climbing the asymptotic giant branch (AGB). During the core-He-burning phase, thermohaline instability develops and slightly changes the surface abundances of Li again and $^{12}$C/$^{13}$C (see Fig. \ref{thepredic}). The second dredge-up leads to a decrease again in the Li abundance and, slightly, in $^{12}$C/$^{13}$C. As shown by CL10, thermohaline mixing is responsible for the Li-enrichment shown during the thermal pulses AGB phase, at T$_{\rm{eff}}\sim$ 3000 K on Fig. \ref{thepredic} (red line). It can explain the Li-enrichment observed during the TP-AGB phase.

\subsection{The variation of the asymptotic period spacing during stellar evolution}

The period spacing of the dipolar gravity modes, $\Delta\Pi$($\ell$=1), can be determined with the asymptotic relation \citep{Tassoul80}.  
\begin{equation}
\Delta\Pi (\ell=1) =\frac{\sqrt{2} \pi^{2}} {\int_{r_{1}}^{r_{2}} N \frac{dr}{r}}
\end{equation} 

where N is the  Brunt-V\"ais\"al\"a frequency and r$_{1}$ and r$_{2}$ define the domain (in radius) where g modes are trapped. Within this region, the mode frequency satisfies the conditions
\begin{equation}
\omega^2 < N^2
\end{equation}
and
\begin{equation}
\omega^2 < S_{\ell}^{2}=\frac{\ell (\ell+1)c_{s}^2}{r^2}
\end{equation}
where $S_{l}$ is the Lamb frequency, and c$_{s}$ the sound velocity.

\begin{figure} 
	\centering
		\includegraphics[angle=0,width=0.5\textwidth, clip=true,trim=1cm 6.5cm 1cm 4cm]{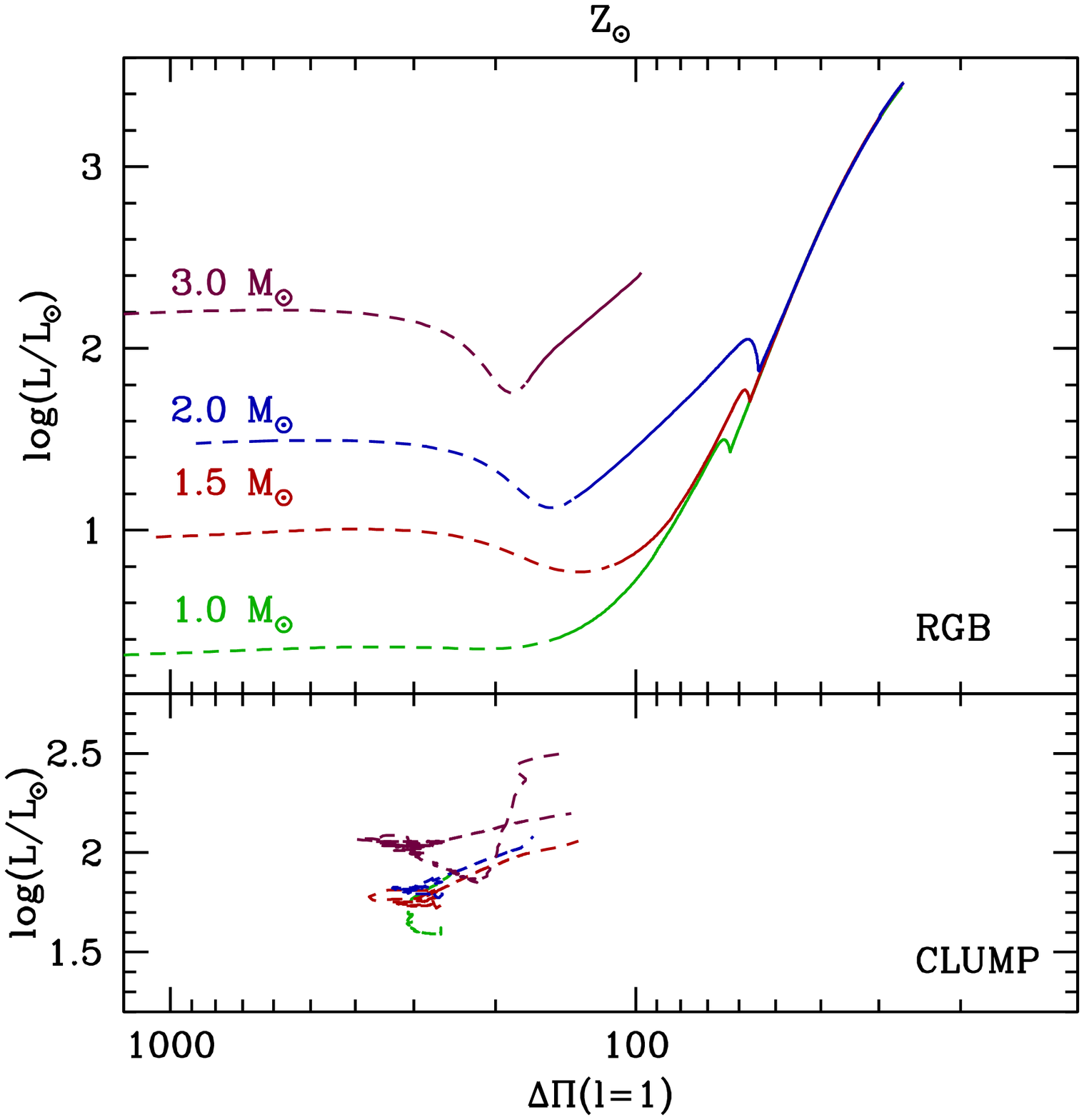}
			  \caption{The stellar luminosity as a function of the asymptotic period spacing of g-modes for standard models at solar metallicity and for different stellar masses, as indicated. Sub-giant and red giant phases are labelled by dashed, and solid lines on the upper panel, which shows the red giant branch. The lower panel shows the He-burning phase.} 
	\label{deltaPipredic}
\end{figure}

This quantity gives us information on the stellar structure, and on the stellar core \citep[][ L12a]{Mosser12a, Montalban13}. Thus with the period spacing of g-modes, it is possible to distinguish between stars with the same luminosity but in different evolutionary phases. A difference in the density profile, and the presence of a convective core during the core-He-burning phase, changes the value of  $\Delta\Pi$($\ell$=1). At the same luminosity, a RGB star has a lower $\Delta\Pi$($\ell$=1) than a clump star ($\Delta\Pi$($\ell$=1) $\sim$ 60 and 300 s respectively, e.g. \citet{Mosser14,Montalban10}). 
Figure \ref{deltaPipredic} depicts $\Delta\Pi$($\ell$=1) for standard models of different initial stellar masses at solar metallicity.  The density profile is different when the stellar mass increases, and, as a consequence, has an impact on the value $\Delta\Pi$($\ell$=1) for a given evolutionary phase (see Fig. \ref{deltaPipredic}). 

As we discuss in Sect. \ref{theo}, rotation-induced mixing has an impact on the stellar structure during the main sequence, and leads to a more massive helium core at the turnoff than in the standard case. This results in a shift of tracks toward higher luminosities throughout their evolution \citep[e.g., L12a,][]{Eggenberger10,Ekstrom12, MaMe00}.  
The effect of increasing the mass of the He core at the TO is to change  the time spent in the sub-giant phase, as well as  the period spacing (for stars with masses ~2.0 M$_{\odot}$) when the star will reach the phase of central He-burning (as explained in Montalban et al. 2013). \citet{Montalban13} already investigated  the effect of main sequence overshooting on the period spacing of intermediate-mass stars during the core-He burning, and also the effect of overshooting during the core-He burning for low-mass stars. A detailed study of the impact of rotation is in progress (Lagarde et al, in prep.).

As discussed in \S\ref{theo}, thermohaline mixing is efficient only after the bump luminosity on the RGB \citep{ChaZah07a, ChaLag10}. Beyond this point, the double-diffusive instability develops in a very thin region located between the hydrogen-burning shell and the convective envelope, and has a negligible effect on the stellar structure. It does not modify the tracks in the $\log(L)$ versus $\Delta\Pi$ diagram.

%%%%%%%%%%%%%%%%%%%%%%%%%%%%%%%%%%%%%%%%%%%%%%%%%%%%%%%%%
%%%%%%%%  Comparaison obs %%%%%%%%%%%%%%%%%%%%%%%%%%%%%%%%%%%%%%%%
%%%%%%%%%%%%%%%%%%%%%%%%%%%%%%%%%%%%%%%%%%%%%%%%%%%%%%%%%%

\section{Comparison with spectroscopic observations of CoRoT red-giants targets}
\label{compa}

We now compare the theoretical predictions of our models with respect to observations of lithium and carbon isotopic ratio in red-giants target.

\subsection{Lithium}

\begin{figure*}
	\centering
		\includegraphics[angle=0,width=0.43\textwidth, clip=true,trim = 1cm 6.5cm 1cm 4cm]{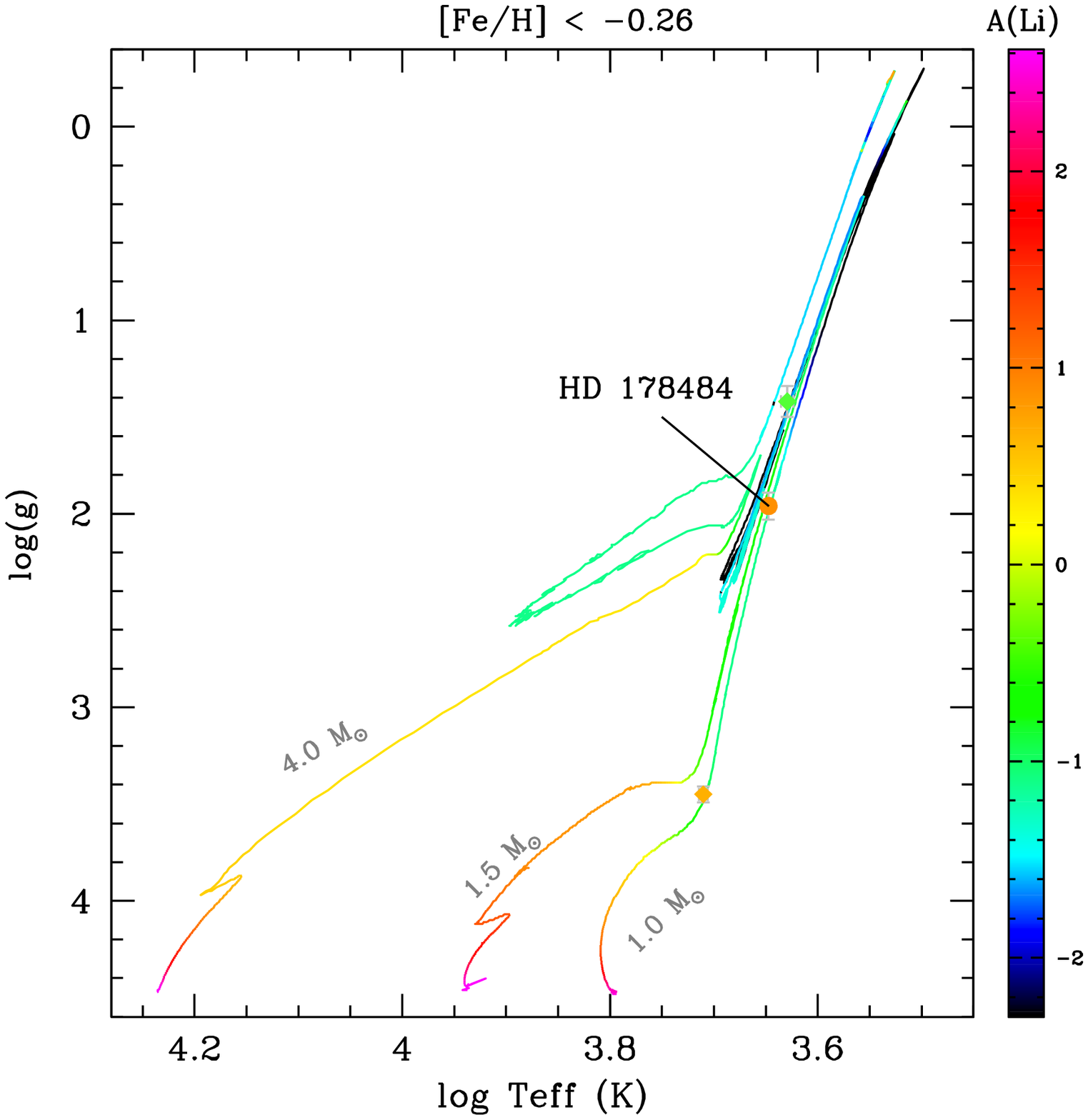}
		\includegraphics[angle=0,width=0.43\textwidth, clip=true,trim = 1cm 6.5cm 1cm 4cm]{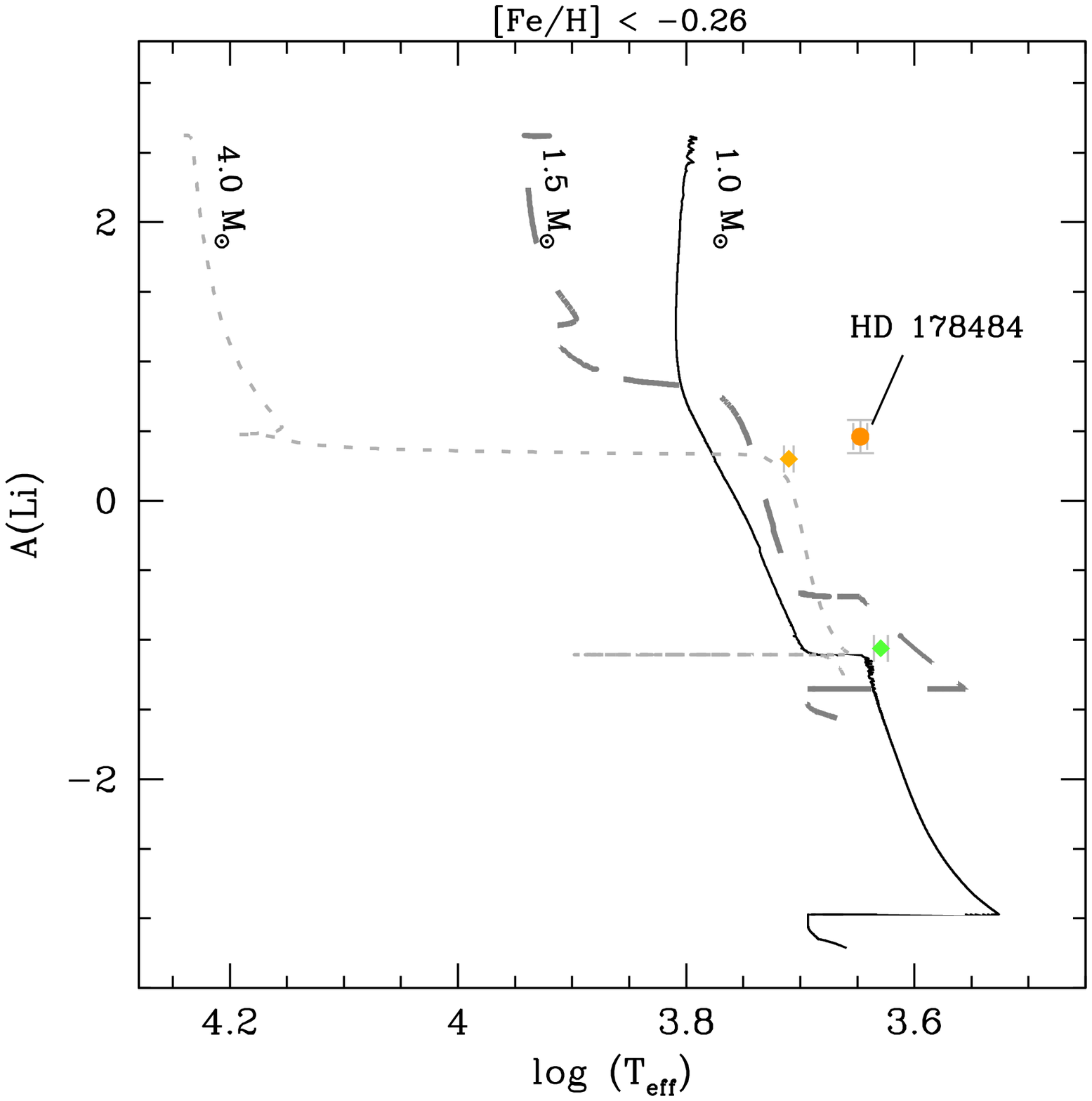}\\
		\includegraphics[angle=0,width=0.43\textwidth, clip=true,trim =1cm 6.5cm 1cm 4cm]{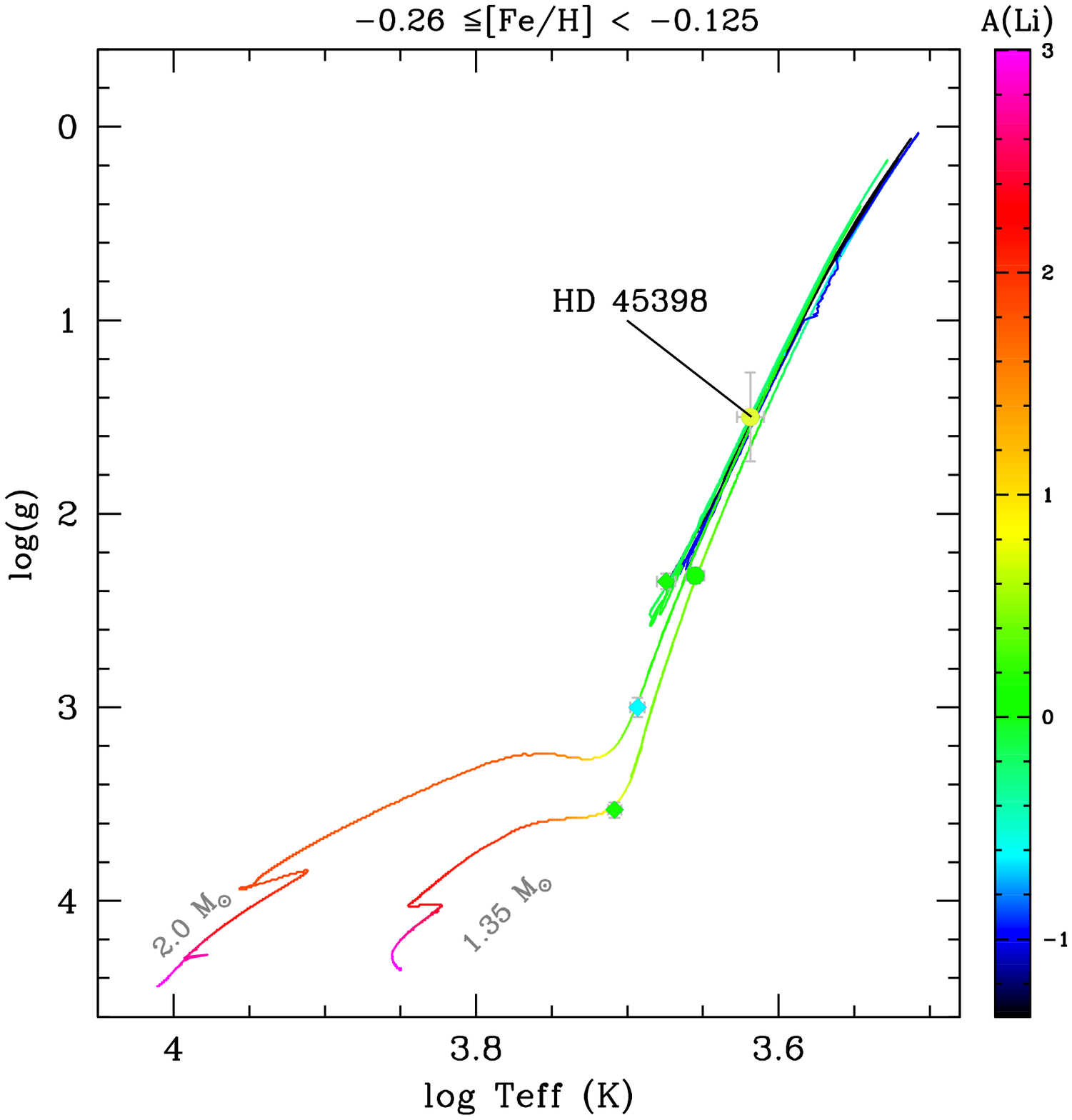}
		\includegraphics[angle=0,width=0.43\textwidth, clip=true,trim = 1cm 6.5cm 1cm 4cm]{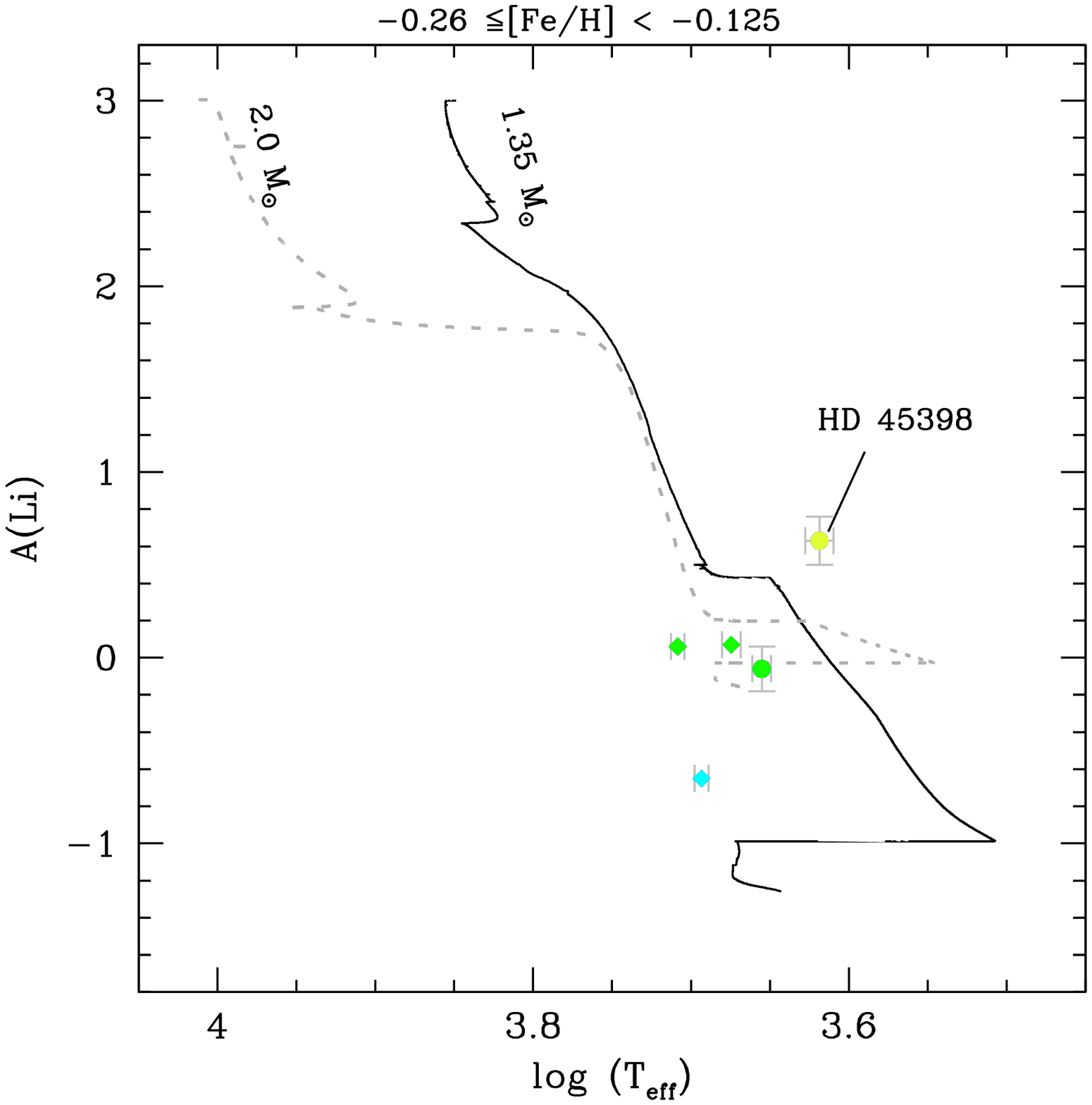}\\
		\includegraphics[angle=0,width=0.43\textwidth, clip=true,trim = 1cm 6.5cm 1cm 4cm]{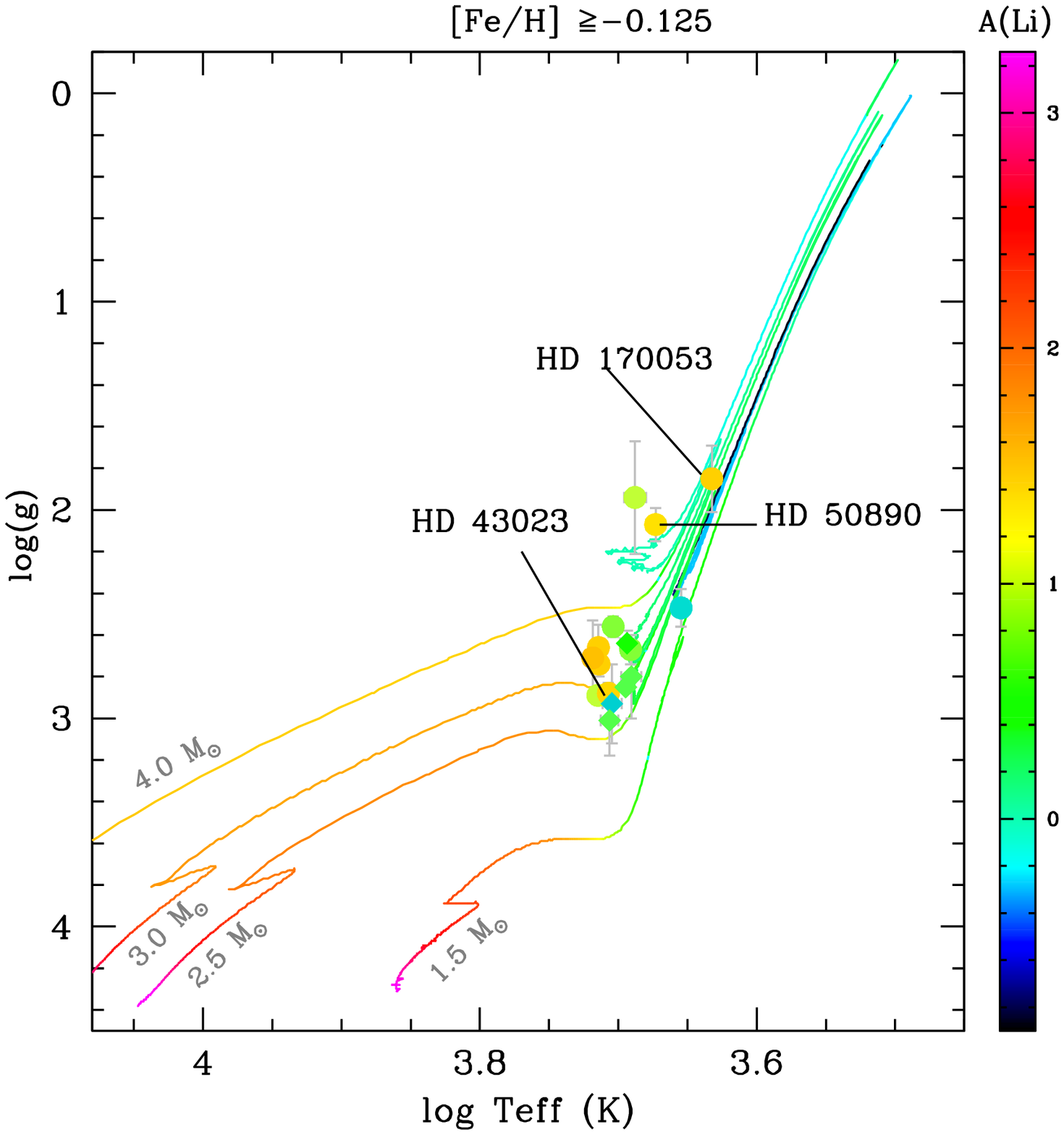}
		\includegraphics[angle=0,width=0.43\textwidth, clip=true,trim = 1cm 6.5cm 1cm 4cm]{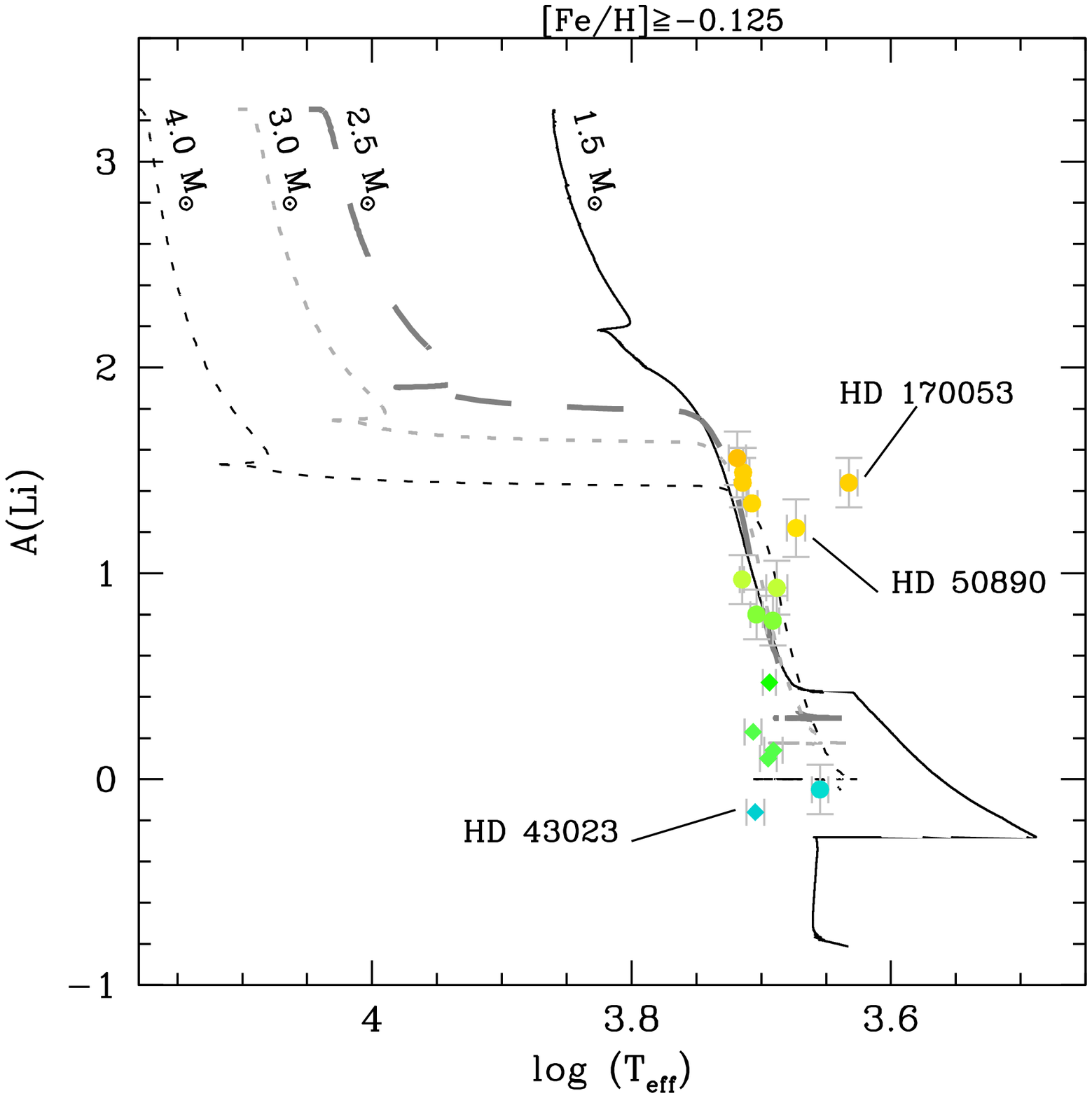}
		  \caption{\textit{Left panels: }Color-coded HR diagram for different stellar masses. The color code represents the values of A(Li) at the stellar surface. \textit{Right panels: } The evolution of surface lithium abundance (from the ZAMS to the end of the He-burning phase) as a function of effective temperature. Circles and diamonds denote, respectively, Li detections and upper limits for stars with [Fe/H]$<$-0.26 (top panels), -0.26$\leq$[Fe/H]$<$-0.125 (middle panels), and [Fe/H]$\geq$-0.125 (bottom panels). Error bars are shown for all stars.}
			 
	\label{Lisol}
\end{figure*}

Figure \ref{Lisol} presents theoretical lithium evolution 
for different masses, at three metallicities ([Fe/H]= -0.56, -0.25, and 0). The models include rotation with an initial velocity $\sim$30\% of critical velocity, which is typical of the observed rotation in the main-sequence stars in young open clusters (see $\S$\ref{TH}, and L12). Theoretical tracks are compared to non-LTE Li observations in three metallicity ranges: [Fe/H]$\geq-0.125$, $-0.26\leq\rm[{\rm Fe} /{\rm H}]<-0.125$, and [Fe/H]$<-0.26$. Our sample does not include Li-rich giants with A(Li)$>$2.5. The stellar surface gravity is taken from seismology for the stars observed by CoRoT. Otherwise we use stellar gravity deduced from spectroscopy.  \\

In most cases, the theoretical and observed lithium abundances are compatible (within the error bars). However, few cases should be discussed in detail:\\ 

\begin{itemize}

 \item \textbf{HD 178484} (on top panels of Fig.\ref{Lisol})  has an observed surface lithium abundance A(Li)=0.46$\pm$0.12 higher than predicted by models (Fig.\ref{Lisol}). As discussed by \citet{Smiljanic10} and in Sect. 4, the  lithium post-dredge up abundance is dependent on the initial values of the stellar rotational velocities (see Fig. \ref{thLi2p0}).  If we were to adopt a lower initial value for the rotational velocity then better agreement would be seen between observation and theoretical models. This star could be a red-giant star close to the bump luminosity with a low initial velocity. \\ 
\item For the same reason, the surface abundance of lithium in \textbf{HD 45398} (on middle panels of Fig. \ref{Lisol} with  A(Li)=0.63$\pm$0.13, no seismic information) could be explained by a low initial velocity.  We believe that this star ascends the red-giant branch.  \\

\item Similarly to the two stars discussed before, \textbf{HD 43023} (on bottom panels of Fig. \ref{Lisol} with A(Li)<-0.16, no seismic information) is a red-giant star beginning to climb the red-giant branch this time with a higher value of the initial velocity than the model shown because of the lower than expected Lithium abundance. \\
 
 \item \textbf{HD 50890}
 (with A(Li)=1.22, on bottom panel of Fig. \ref{Lisol}) has been studied by \citet{Baudin12}. They concluded, by modelling seismic properties, that it is a core He-burning star with mass in the range between 3 and 5 M$_{\odot}$. Its location in the log$g$-T$_{\rm{eff}}$ diagram yields a similar stellar mass. However, due to the large uncertainties on Li abundance, we cannot confirm the evolutionary status. \\
 
  \item All members of \textbf{NGC 6633}, will be discussed in detail in Sect. \ref{ngc6633part}. \\

\end{itemize}
\subsection{Carbon isotopic ratio}

\begin{figure}
	\centering
		\includegraphics[angle=0,width=0.5\textwidth, clip=true,trim = 1.5cm 7cm 1cm 4cm]{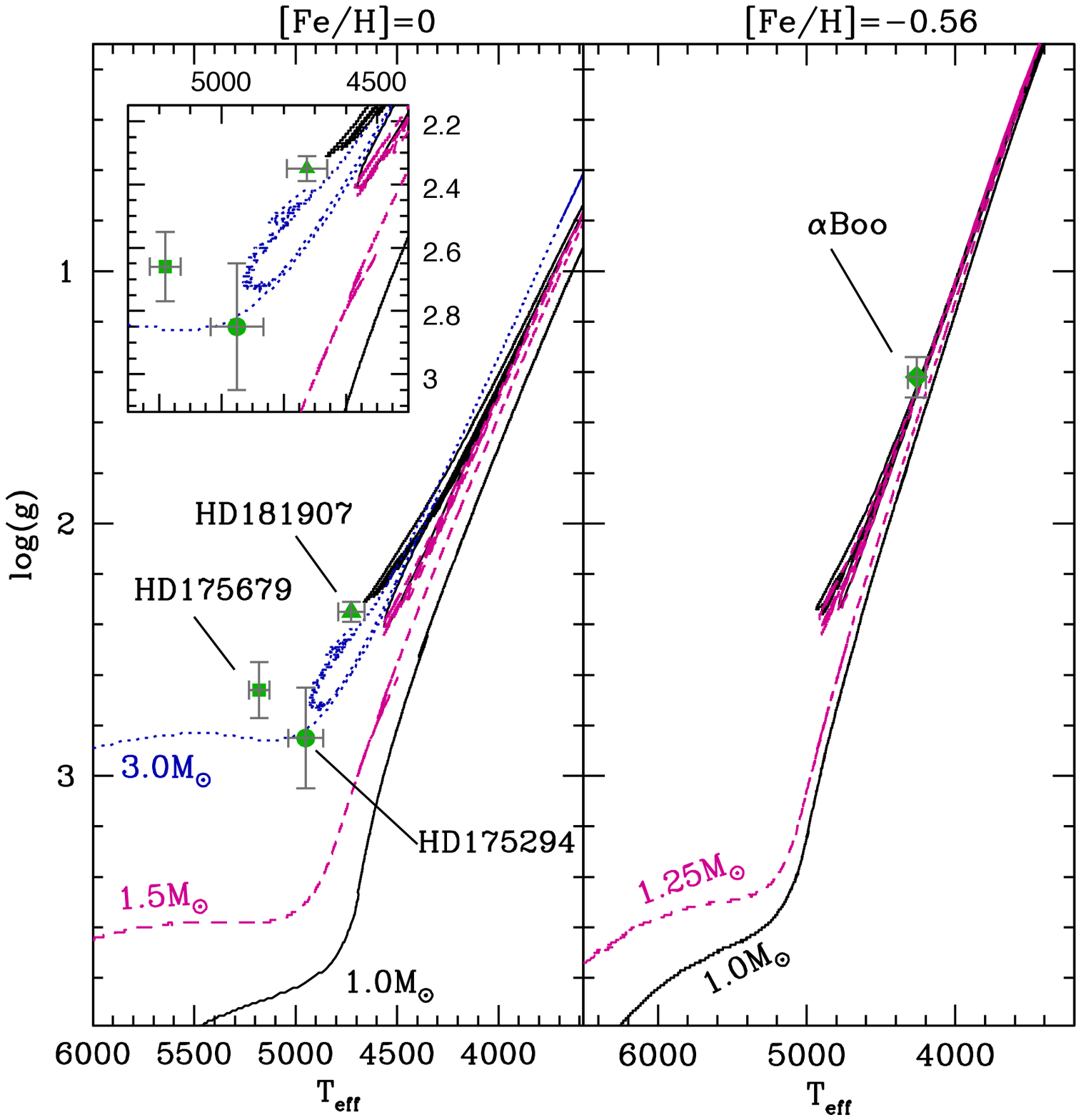}
			  \caption{Theoretical evolutionary tracks plotted in log$g$-T$_{\rm{eff}}$ diagram (from the main sequence up to the early-AGB) computed with thermohaline instability and rotation-induced mixing at solar metallicity (left panel) and at [Fe/H]=-0.56 (right panel). Different line styles correspond to different stellar masses. Position of the considered set of stars for which $^{12}$C/$^{13}$C has been measured are represented by a circle for HD 175294, a square for HD 175679, a triangle for HD 181907, and a diamond for $\alpha$ Boo;  They are segregated according to their metallicity.}
	\label{logg_c1213_obs}
\end{figure}

\begin{figure} 
	\centering
		\includegraphics[angle=0,width=0.5\textwidth, clip=true,trim = 1.5cm 7cm 1cm 4cm]{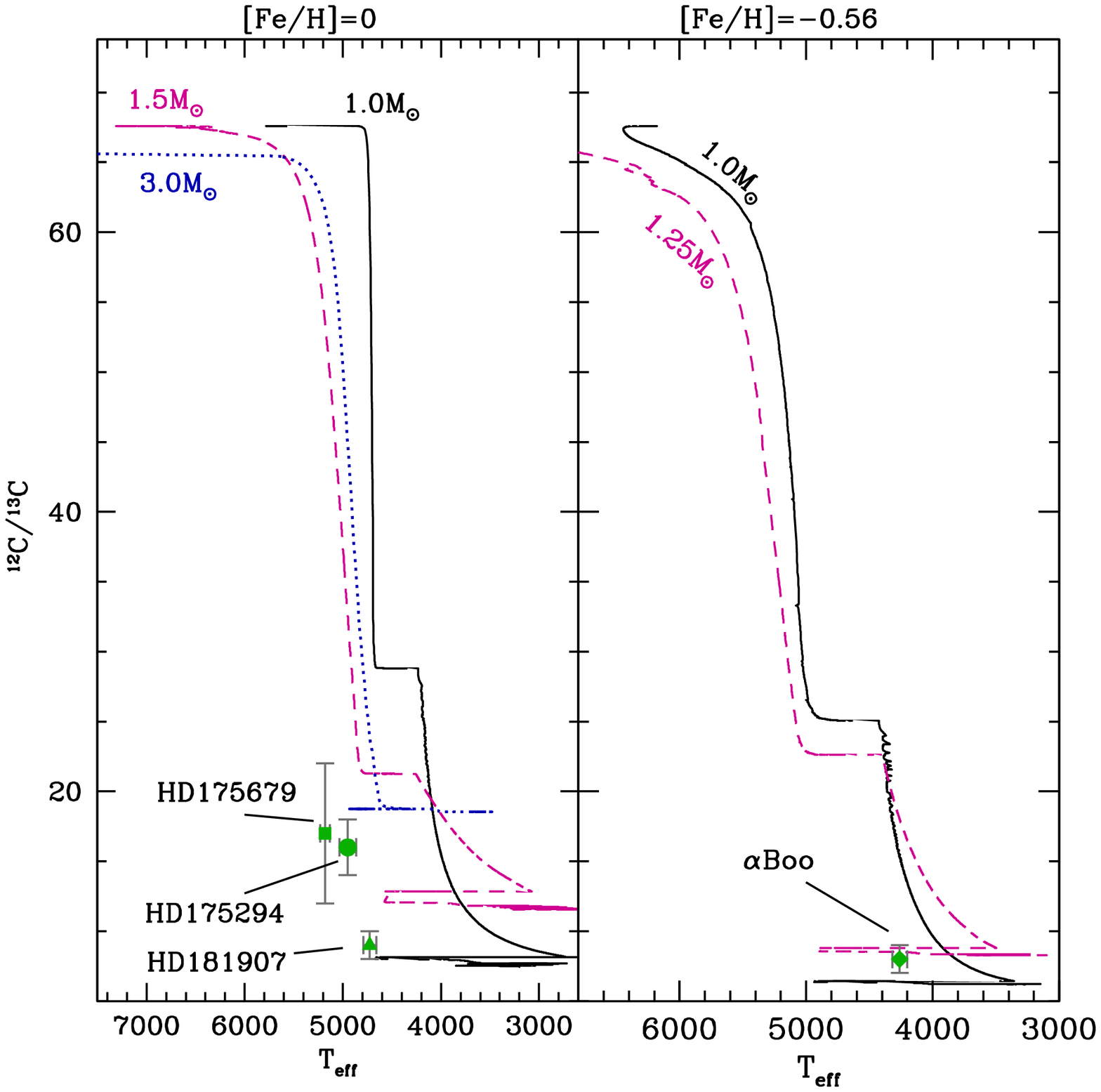}
			  \caption{$^{12}$C/$^{13}$C data in our red-giant stars that are segregated according to their metallicity (left and right panels include respectively sample stars with metallicity close to solar and [Fe/H]=-0.56). Same symbols as in Fig \ref{logg_c1213_obs}. Theoretical $^{12}$C/$^{13}$C surface abundance is shown from the ZAMS up to the TP-AGB. Various lines correspond to predictions of stellar models of different masses including effects of rotation-induced mixing (with an initial V/V$_{\rm{crit}}$=0.30) and thermohaline instability.}
	\label{c1213_obs}
\end{figure}

\begin{table}
  \caption{Chemical properties of stars with $^{12}$C/$^{13}$C data and NGC 6633 members}
                      \label{dataC}  

	\scalebox{0.78}{
         \begin{threeparttable}
         \centering   

\begin{tabular}{|c || c | c | c || c | c |}
    \hline
     Name & T$_{\rm{eff}}$ & log(g) & [Fe/H] & $^{12}$C/$^{13}$C  & A(Li)$_{\rm{NLTE}}$   \\ 
     & (K) & & & &  \\
     \hline
     HD181907 & 4725 $\pm$ 65 & 2.35 $\pm$ 0.04 & -0.15 $\pm$0.12 & 9 $\pm$ 1& $<$ 0.07  \\
    HD175679 & 5180 $\pm$ 50&2.66 $\pm$ 0.11&0.02 $\pm$0.10 & 17$\pm$ 5& 1.44 $\pm$ 0.12   \\
    HD175294$^{[3]}$ & 4950 $\pm$ 85 & 2.85 $\pm$ 0.2& 0.25 $\pm$ 0.12 & 16$\pm$2 & $<$ 0.1 \\
      $\alpha$ Boo& 4260 $\pm$ 60&1.42 $\pm$ 0.08 & -0.69 $\pm$ 0.11 & 8$\pm$1 & $<$ 1.06  \\
   \hline
     \hline
      HD170053 $^{[2]}$ & 4290 $\pm$ 65 & 1.85 $\pm$ 0.16 & -0.03 $\pm$0.12 & 18 $\pm$ 8 $^{[1]}$& 1.44$\pm$ 0.12  \\
     HD170174 $^{[2]}$ & 5055 $\pm$ 55 & 2.56 $\pm$ 0.05 & -0.07 $\pm$0.10 & 21 $\pm$ 7 $^{[1]}$ & 0.8 $\pm$ 0.12  \\
     HD170231$^{[2]}$  & 5175 $\pm$ 55 & 2.74 $\pm$ 0.06 & -0.03 $\pm$ 0.10 & - & 1.49 $\pm$ 0.12  \\
   \hline
   
 \end{tabular}
 
        \begin{tablenotes}
        \item[1] from \citet{Smiljanic09}
        \item[2] NGC 6633 members
        
       	\end{tablenotes}
	
	\end{threeparttable}
	}

\end{table}

As discussed in Sect. \ref{obs}, M14 investigated the carbon isotopic ratio in four stars. In this section, we propose to discuss in detail these cases by comparing seismic and spectroscopic properties. In order to locate these stars on the evolutionary tracks, Fig.\ref{logg_c1213_obs} presents the theoretical logg as a function of effective temperature for different masses and at two metallicities. In addition, the theoretical $^{12}$C/$^{13}$C surface evolution with T$_{\rm{eff}}$ for different stellar masses at two metallicities are shown in Figure \ref{c1213_obs}, and compared with the observations. \\

These four stars present low carbon isotopic ratios which are not predicted by standard models. However, effects of rotation and more significantly, thermohaline mixing can account for such a decrease of $^{12}$C/$^{13}$C surface abundances and reproduce very nicely the spectroscopic observations. The stellar mass, and metallicity have an impact on the effects of thermohaline instability and rotation-induced mixing on stellar surface abundances \citep[for more details see ][and Lagarde \& Charbonnel in prep.]{ChaLag10}. \\

The upper limits of Li abundances for \textbf{HD 175294} (circle) and \textbf{Arcturus} ($\alpha$Boo, diamond) do not give additional constraints, although they are consistent with their carbon isotopic ratio. \\
According to its low $^{12}$C/$^{13}$C value (right panel of Fig. \ref{c1213_obs}), \textbf{Arcturus} is an early-AGB star that has already finished the core He-burning phase, with a stellar mass between 1.0 and 1.25 M$_{\odot}$. 
Its position on the evolutionary tracks (right panel Fig. \ref{logg_c1213_obs}) confirms its early-AGB status. 
The seismic properties were derived by \citet{Kallinger10a}. As shown by \citet{Miglio12}, using the interferometric radius does not significantly change the stellar mass. Indeed, as shown in Fig. \ref{sample_M} and in Fig. 3 of \citet{Miglio12}, the stellar mass of Arcturus is between 0.6 and 0.9 M$_{\odot}$. The seismic mass is significantly below that inferred from the models. However as Arcturus is believed to be an early-AGB, we can expect it to have experienced high mass loss. Indeed, using Reimers formula in our models, we find that a star with an initial mass between 1.0 and 1.2 M$_{\odot}$ at [Fe/H]=$-0.56$ has a mass between 0.9 and 1.17 M$_{\odot}$  on the early-AGB.\\ 

\textbf{HD 175294}, was initially proposed as potential target for CoRoT, but was not observed. Consequently, we have access only to the spectroscopic surface gravity which is more uncertain. From spectroscopic point of view, this star is likely to be a core He-burning star with a stellar mass around 3.0 M$_{\odot}$. \\

\textbf{HD 175679} (square on Figs. \ref{logg_c1213_obs} and \ref{c1213_obs})  has seismic data with a very large uncertainties that prevent an accurate estimation of a seismic mass and radius.  The Li abundance is in agreement with that of a star with a mass range of 3 to 4 M$_{\odot}$ located in the Hertzsprung-gap or possibly at the base of the RGB.  Although the carbon isotopic ratio is uncertain, the value is consistent  with this evolutionary state and initial mass. \\ 

The properties of the red giant \textbf{HD181907} (HR 7349, triangle on Figs. \ref{logg_c1213_obs} and \ref{c1213_obs}) has widely been discussed in the literature. Using solely the seismic observations by \citet{Carrier10}, \citet{Miglio10} deduce a mass of about $1.2\,M_{\odot}$, which is in good agreement with the value we deduced from spectroscopic constraints (see Figs.  \ref{logg_c1213_obs} and \ref{c1213_obs}), as well as with stellar mass and radius deduced from the Hipparcos parallax (Fig. \ref{sample_M}). 

An observation of the small frequency separation between $\ell=0$ and $\ell=1$ modes, $ \delta \nu_{\rm{01}}$, allowed \citet{Montalban10} to suggest that this star is ascending or descending the RGB. 
However, the value of the small frequency separation between $\ell=0$ and $\ell=2$ modes, $ \delta \nu_{\rm{02}}$, \citep{Carrier10} seems to be too large for a RGB star \citep{Montalban12b}. In addition its low surface carbon isotopic ratio and Li abundances indicate that this star would be better explained by a model in the core-He burning phase (Fig. \ref{c1213_obs}).\\

Further seismic analysis can be used to distinguish between the RGB and core-He burning phase. Measurement of the period spacing indicates that a value of around 285\,s can fit the modes. This solution is not unique due to the limited number of mixed modes observed with a 5-month run. However we can rule out a period spacing of about 60\,s typical of an RGB star. We also applied the method of \citet{Kallinger12} to estimate the evolutionary status, but again the frequency resolution makes it difficult to derive a firm conclusion. Further confirmation on the evolutionary state is derived from a comparison of the oscillation spectrum of HD 181907 with spectra of \textit{Kepler} red giants with similar large separations and $\nu_{\rm{max}}$ values, and with identified evolutionary stages \citep{Mosser14}. In that respect, the spectrum of HD 181907 looks like a red-clump star spectrum, with significant power in the gravity-dominated mixed modes \citep{Grosjean14}. We therefore believe this star to be a core-He burning star, based on both spectrometric and asteroseismic arguments. 
Nevertheless, it would be interesting to develop more detailed models of this star, by computing theoretical oscillation frequencies directly from stellar models including the effects of, e.g., thermohaline instability and rotation-induced mixing. \\

\begin{figure*} 
	\centering
		\includegraphics[angle=0,width=0.49\textwidth, clip=true,trim=1cm 6.5cm 1cm 4cm]{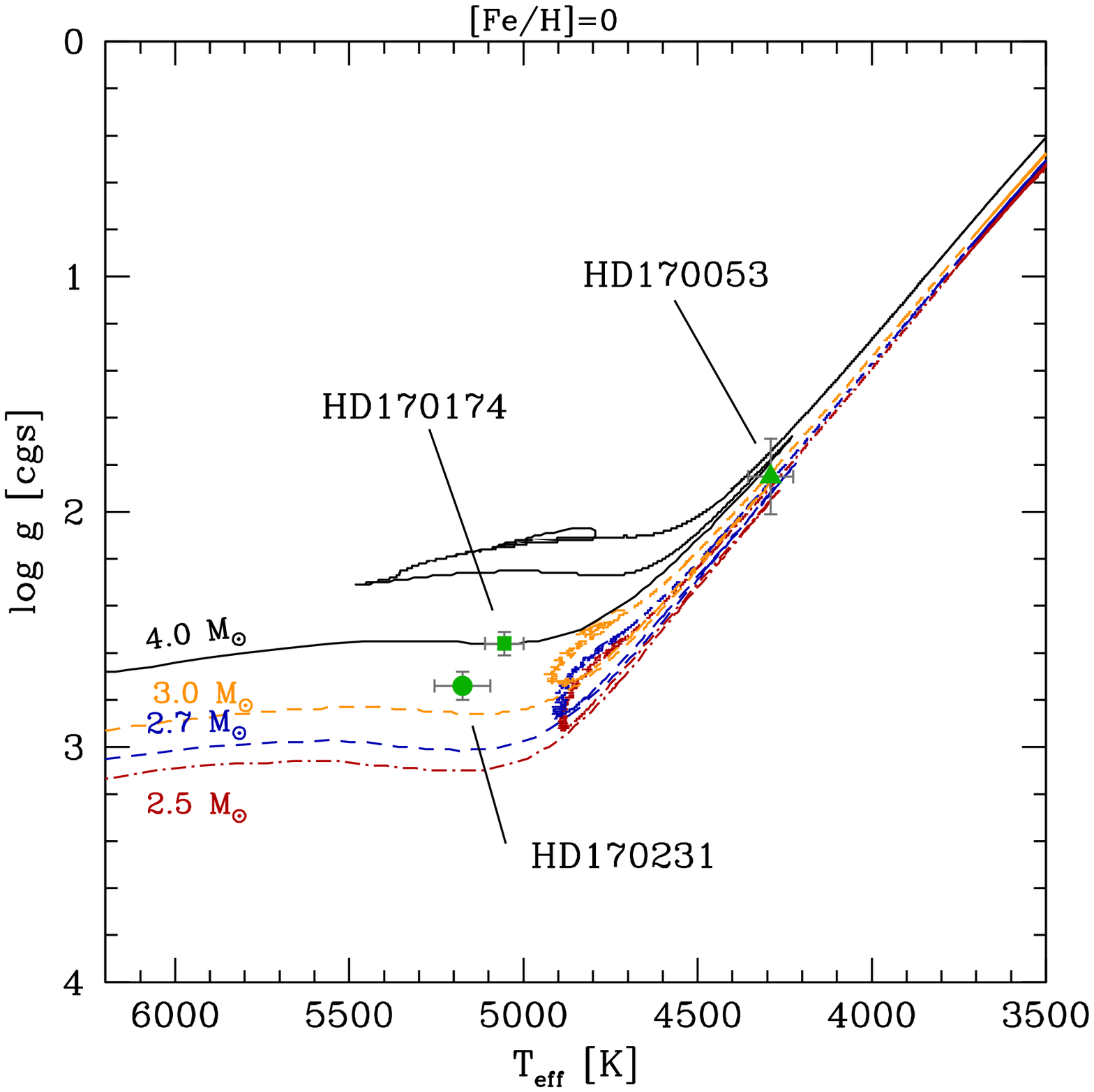}
		\includegraphics[angle=0,width=0.49\textwidth, clip=true,trim=1cm 6.5cm 1cm 4cm]{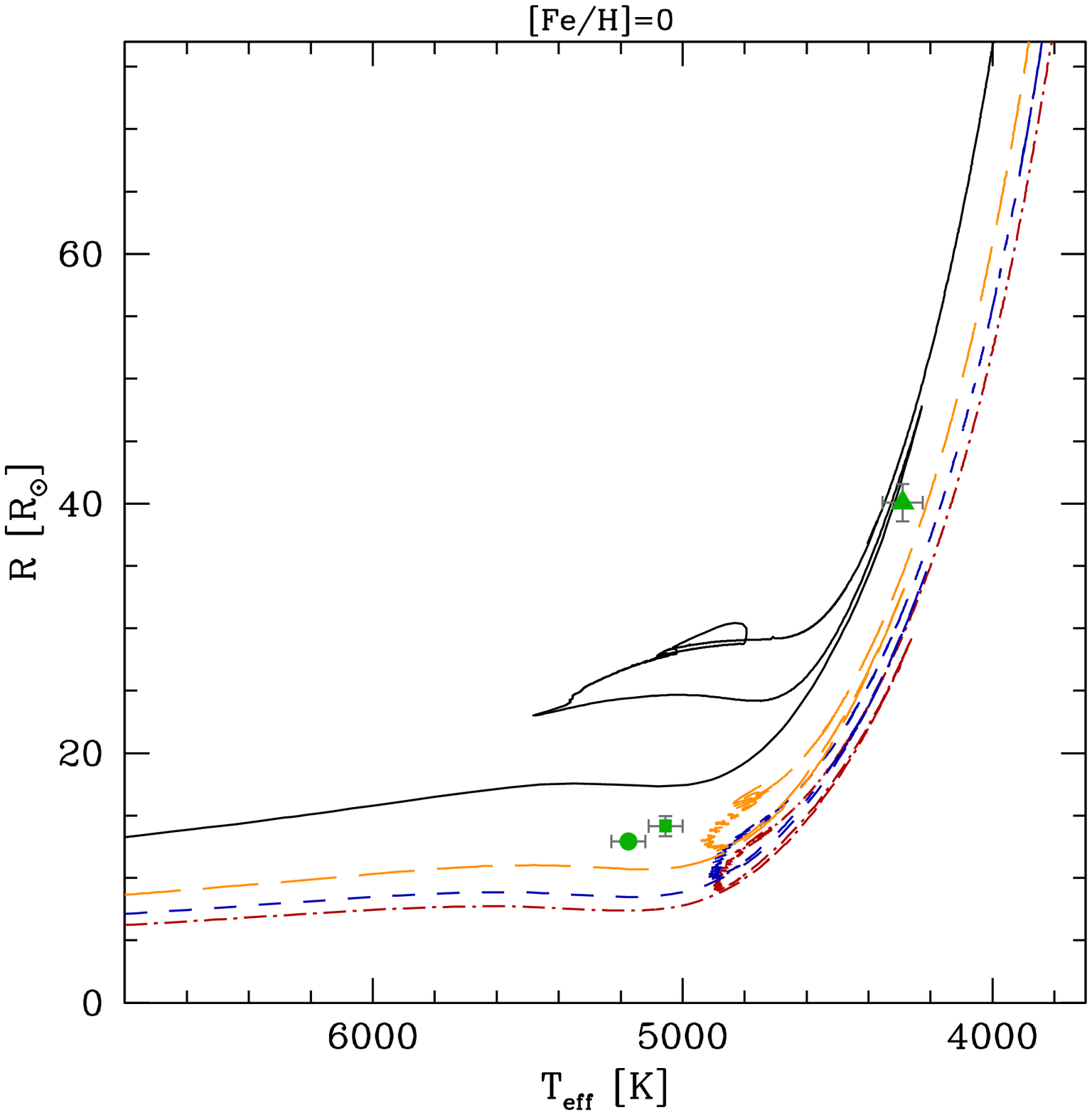}
			  \caption{ Theoretical evolutionary tracks plotted in logg-Teff diagram (\textit{left panel}) and stellar radius (\textit{right panel}) (from the main sequence up to the early-AGB) computed with thermohaline instability and rotation-induced mixing at solar metallicity for 4.0 M$_{\odot}$ (V$_{\rm{ZAMS}}$=144 km/s, solid black line), 3.0 M$_{\odot}$ (V$_{\rm{ZAMS}}$= 136 km/s, orange long dashed line), 2.7 M$_{\odot}$ (V$_{\rm{ZAMS}}$=110 km/s, blue dashed line), and 2.5 M$_{\odot}$ (V$_{\rm{ZAMS}}$=110 km/s, red dashed line). Cluster members discussed in this study are indicated by green circle (HD170231), square (HD170174), and triangle (HD170053).} 
	\label{NGC6633_logg}
\end{figure*}

\subsection{NGC 6633}
\label{ngc6633part}

 \begin{figure*}
	\centering
		\includegraphics[angle=0,width=0.49\textwidth, clip=true,trim=1cm 6.5cm 1cm 4cm]{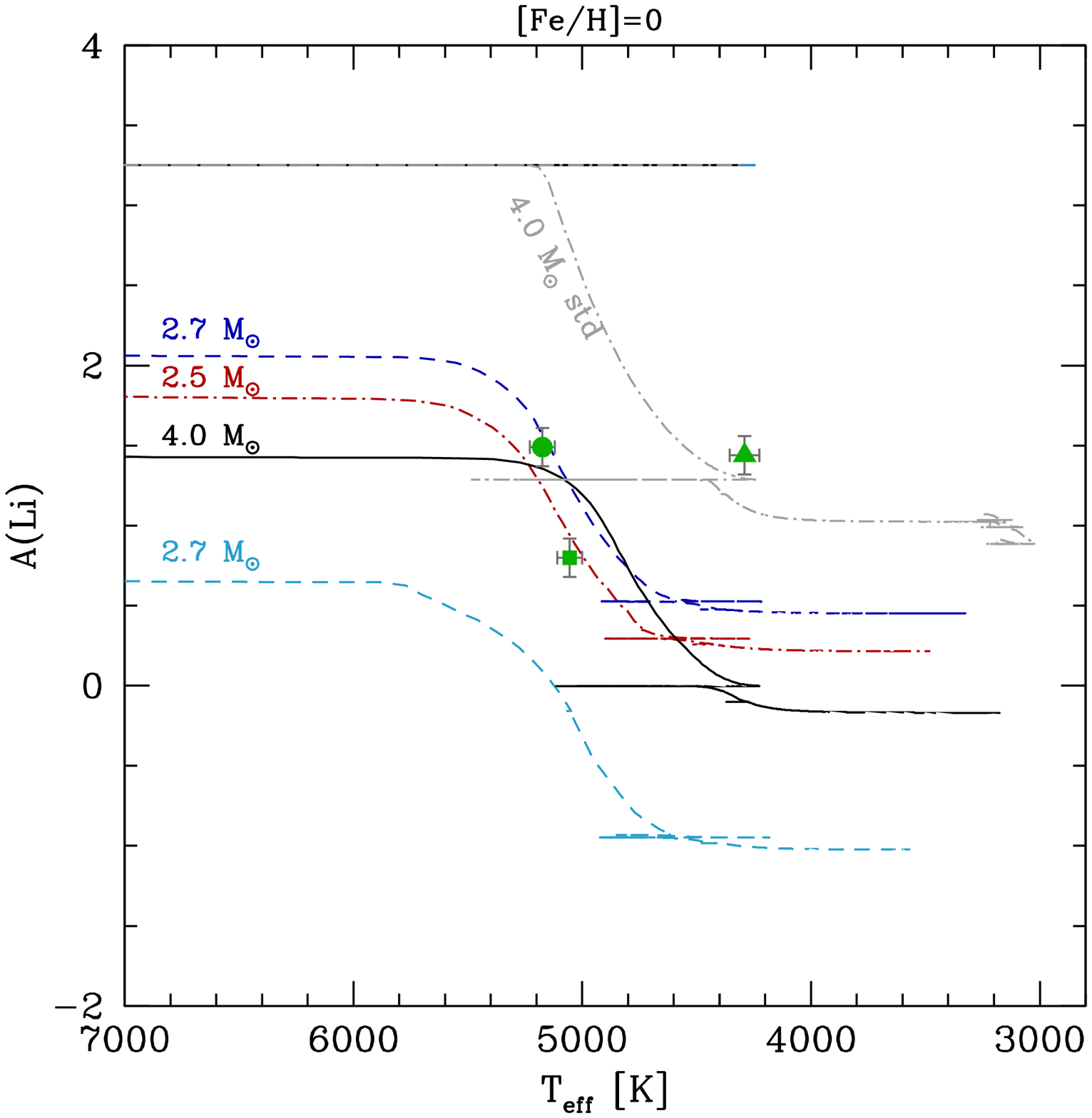}
		\includegraphics[angle=0,width=0.49\textwidth, clip=true,trim=1cm 6.5cm 1cm 4cm]{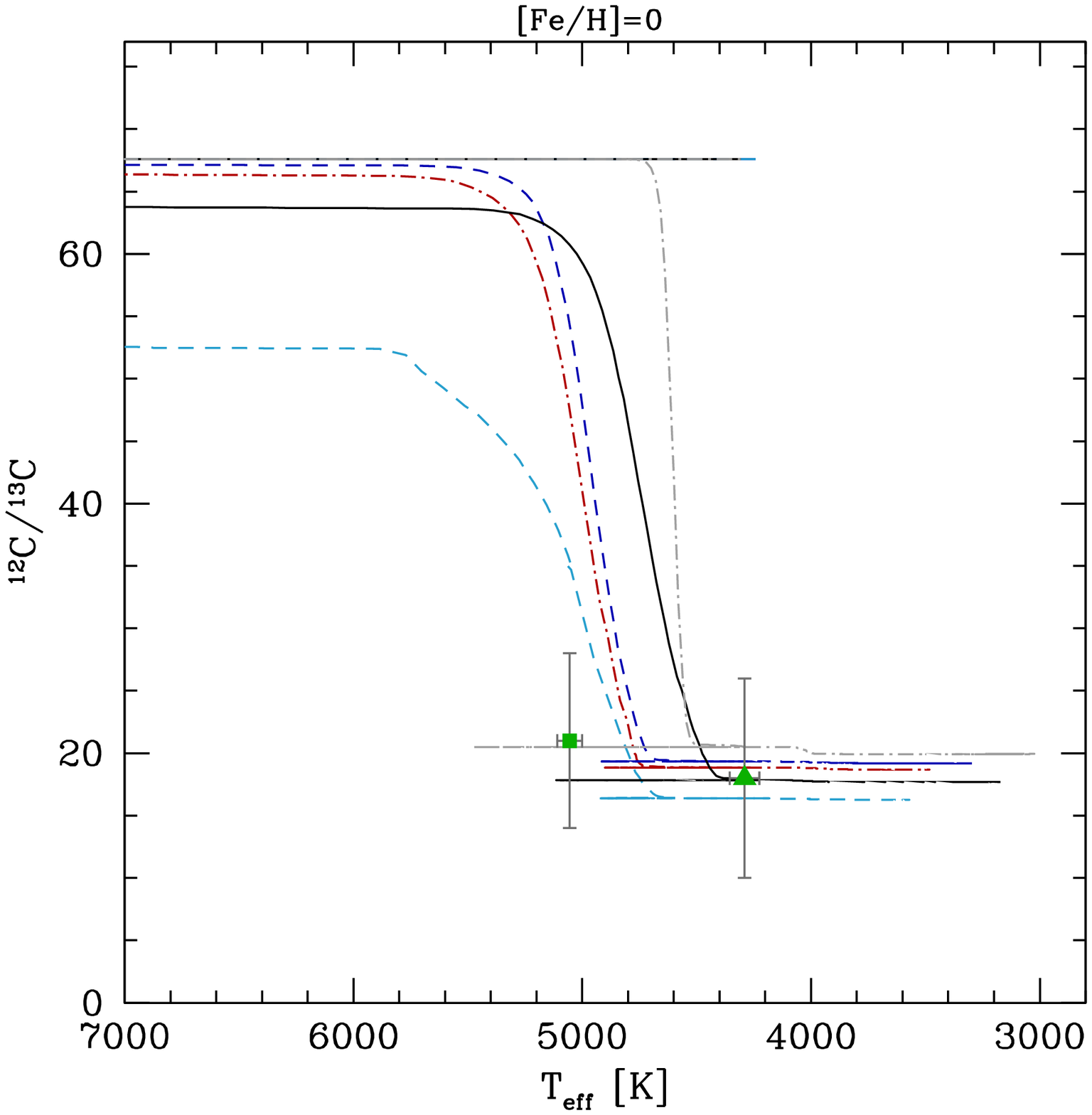}
			  \caption{Theoretical evolution of lithium A(Li) (left panel) and the carbon isotopic ratio (right panel) as a function of effective temperature (from the main sequence up to the early-AGB) computed with thermohaline instability and rotation-induced mixing at solar metallicity for 4.0 M$_{\odot}$ (V$_{\rm{ZAMS}}$=144 km/s, solid black line), 2.7 M$_{\odot}$ (V$_{\rm{ZAMS}}$=110 km/s and 250 km/s, blue and light blue dashed lines respectively), and 2.5 M$_{\odot}$ (V$_{\rm{ZAMS}}$=110 km-s$^{-1}$, red dashed line). Model of 4.0M$_{\odot}$ following standard evolution theory is also indicated by dashed-dotted line. Lithium detection and $^{12}$C/$^{13}$C determination for cluster members are indicated by green symbols.}
	\label{NGC6633_LiC}
\end{figure*}

Since stars belonging to a cluster have been formed together, we can assume that they have the same age, distance, and metallicity. Spectroscopic and asteroseismic data for these stars represent a unique opportunity to improve the constraints on mixing inside red-giant stars. CoRoT has detected solar-like oscillations in three members of open cluster NGC 6633 \citep[HD 170053, HD 170174, HD 170231; ][]{Barban14,Poretti14}. In addition, M14 and \citet{Smiljanic09} present spectroscopic studies of these stars. The lithium abundances are taken from M14 and the carbon isotopic ratio for two stars from \citet{Smiljanic09}. The age of the cluster determined by isochrone fitting in \citet{Smiljanic09} (t=4.5.10$^{8}$yrs) implies that stars in the He-core-burning stage have 2.8 $\lesssim$ M/M$_{\odot}$ $\lesssim$ 3.0, which is compatible with the stellar mass determined with asteroseismology.\\

\citet{Molenda14} very recently presented a study of 5 stars in NGC 6811 observed by \textit{Kepler}, NGC 6633 is the only cluster for which spectroscopic probes of extra-mixing on RGB (Li and $^{12}$C/$^{13}$C) and asteroseismic properties are available. All spectroscopic and asteroseismic properties for members of NGC 6633 are listed in table \ref{dataC}.\\

Figure \ref{NGC6633_logg} presents observations in $\log \rm g-T_{\rm{eff}}$ diagram and stellar radius vs T$_{\rm{eff}}$. In the same figure, four evolutionary tracks for 2.5, 2.7 and 4.0 M$_{\odot}$ at solar metallicity are shown. These tracks include the effects of rotation and thermohaline mixing. The turnoff mass of NGC 6633 lies in the range 2.4 to 2.7 M$_{\odot}$ \citep{Smiljanic09}. \\

Figure \ref{NGC6633_LiC} presents chemical properties of the cluster members with lithium (left panel) and carbon isotopic ratio (right panel). 
Intermediate-mass stars ignite central helium-burning in a non-degenerate core and at relatively low luminosity on the RGB, well before the HBS reaches the mean molecular weight discontinuity caused by the first dredge-up. Consequently, thermohaline mixing does not take place in those stars. Only rotation has an impact on surface abundances (see right panel on Fig.\ref{NGC6633_LiC}). \\

In the following paragraphs, we discuss each star according to its seismic and spectroscopic properties: 
\begin{itemize}
\item \textbf{HD 170053} (triangle in Figs. \ref{NGC6633_logg} and \ref{NGC6633_LiC}): According to its position in the color-magnitude diagram and using isochrones from \citet{Schaller92}, \citet{Smiljanic09} proposed that this star could be an early-AGB star. This is perfectly in agreement with the carbon isotopic ratio around 18 $\pm$ 8 that they deduced (see right panel of Fig. \ref{NGC6633_LiC}). As discussed in Sect. \ref{theo}, rotation has an impact on stellar structure during the main sequence. This significantly changes the lithium profile at the end of the main sequence and the surface abundances in the sub-giant phase \citep[e.g.,][]{Palacios03, Smiljanic10}. Taken together logg, T$_{\rm{eff}}$, seismic radius and the values of Li and $^{12}$C/$^{13}$C suggest this star is an \textbf{early-AGB star} with low initial velocity (see Fig. \ref{thLi2p0}).\\

\item \textbf{HD 170174} (square in Figs. \ref{NGC6633_logg} and \ref{NGC6633_LiC}): According to seismic properties this star has a stellar mass 2.7 $\pm$ 0.7 M$_{\odot}$, which is slightly lower than mass deduced from its position on the evolutionary tracks. Indeed, from spectroscopic point of view, HD170174 could be more massive ($\sim$3.0 M$_{\odot}$) as a \textbf{He-burning} star with low initial velocity, or a \textbf{red-giant} star at the bottom of RGB. Due to the very short lifetime on the sub-giant branch, the latter possibility is unlikely. Using the seismic stellar mass (2.7 M$_{\odot}$) and considering this star as an He-burning star, the model with an initial velocity $\sim$95 km/s at the ZAMS is the best model to reproduce its chemical properties.\\

\item \textbf{HD 170231} (circle on Figs. \ref{NGC6633_logg} and \ref{NGC6633_LiC}): According to spectroscopic and seismic properties, this star could be a sub-giant or a \textbf{He-burning star}. The low value of $\Delta \nu$ rules out the possibility that this star is a sub-giant. Its seismic properties give a stellar mass around 3.4 M$_{\odot}$ which is in good agreement with its position on evolutionary tracks and its lithium abundance. A determination of the period spacing of g-modes would allow us to distinguish between these two evolutionary options. 

\end{itemize}
 
 \section{Discussion and conclusions}
 \label{conclu}
 
In this paper, we demonstrate the power of the combination of seismic and spectroscopic constraints to improve our understanding of the physical processes and specifically extra-mixing taking place in the interior of red-giant stars.
Indeed, asteroseismology gives us new informations on stellar interiors and accurate estimates of stellar mass, radius and evolutionary state. Spectroscopy provides complementary information about surface chemical properties of stars.\\

This paper significantly advances the study of CoRoT red giants as presented by M14 by adding a comparison with modern stellar models that incorporate rotation and thermohaline mixing. We compare stellar masses and radii determined using various methods. The estimates are in agreement within a standard uncertainty. However, we found relatively large average uncertainties on radii ($\sim$9\%), and masses ($\sim$22\%) due to large uncertainties on seismic properties ($\Delta \nu$ or $\nu_{\rm{max}}$). These values are dominated by the stars observed in short and initial CoRoT runs. Indeed, these uncertainties are significantly lower when considering only stars observed in 150-d long runs and with an apparent visual magnitude brighter than 8 ($\sim$5\% on radius and $\sim$14\% on mass). These statistical uncertainties are  likely to be larger than systematic uncertainties that may affect these relations. \\
The weighted average of the relative difference between Hipparcos and seismic distances (-0.12$\pm$0.03) indicates a possible disagreement. However, the large uncertainty on these two quantities, prevents us from drawing any firm conclusions. \\

We have also compared theoretical and observational behaviors for lithium and carbon isotopic ratio. Stellar models used in this article, include the effects of rotation-induced mixing \citep{Zahn92,MaZa98}, known to change chemical properties of main sequence and sub-giant stars, and thermohaline instability \citep{ChaZah07a,ChaLag10}, known to govern the surface chemical properties of low-mass RGB stars. 
We show that for low-mass stars as Arcturus and HD181907, the low carbon isotopic ratio is well explained by thermohaline instability. On the other hand, for more massive stars it is rotation that is the most efficient transport process for chemical species. Our models at different initial velocities can explain the surface abundances of lithium and $^{12}$C/$^{13}$C. 

\begin{table}
 \caption{Theoretical surface values of carbon isotopic ratio at the end of the first dredge-up and during the He burning phase.}
         
\scalebox{0.86}{
         \begin{threeparttable}
         \centering 

\begin{tabular}{|c | c | c | c | c | c | c |}
    \hline
     Mass &  & V$_{ZAMS}$ & V$_{ZAMS}$/V$_{crit}$ & $^{12}$C/$^{13}$C & $\Delta$($^{12}$C/$^{13}$C )& $\Delta$(V$_{ZAMS}$)\\ 
     (M$_{\odot}$)& & (km/s)  & & & & (km/s)  \\
     \hline
     & & 0 & - & 25.6 &  &\\
    \cline{3-5} 
      & RGB$^{1}$  & 50 & 0.14 & 23.7 & 1 & $\sim$30 \\
    \cline{3-5} 
     & & 80 & 0.22 &21.6 &  & \\
    \cline{3-5} 
     1.25 & & 110 & 0.30 &18.6 &  & \\
     \cline{2-7}
     & & 0 & - &10.5 &  & \\
    \cline{3-5} 
     &He-B$^{2}$   & 50 & 0.14 & 9.5 & 1 & $\sim$110 \\
    \cline{3-5} 
     & & 80 & 0.22& 9.2 & & \\
    \cline{3-5} 
     & & 110 & 0.30 &8.6  & & \\
     \hline
     
     \hline
     &  & 0 & - & 21.8 & &\\
    \cline{3-5} 
     &RGB$^{1}$ & 110 & 0.27 & 19.4 & 1 & $\sim$70 \\
    \cline{3-5} 
     & & 180 & 0.44 & 17.7 & & \\
    \cline{3-5} 
   2.0  & & 250 & 0.61 & 14.8 & & \\
     \cline{2-7}
     &  & 0 & - & 20.3 & & \\
    \cline{3-5} 
     &He-B$^{2}$ & 110 & 0.27 & 16.7 & 1 & $\sim$70 \\
    \cline{3-5} 
     & & 180 & 0.44 & 15.3 & & \\
    \cline{3-5} 
     & & 250 & 0.61 & 13.4 & & \\
     \hline

 \end{tabular}
 \label{c1213vel}
        \begin{tablenotes}
        \item[1] Post-dredge-up values
        \item[2] Central mass fraction of $^4$He $\sim$ 0.5
      	\end{tablenotes}
	\end{threeparttable}
	}
\end{table}

This study could be more quantitative if the seismic and spectroscopic constraints were more accurate. In addition, the small number of stars limits the conclusions. Our study has however identified the key constraints, and their precision, that are needed for a stringent test of our models.
The desirable scenario is the following: to use asteroseismic and spectroscopic constraints to infer stellar masses to 10\% or better, the evolutionary state (RGB vs. core-He burning), and photospheric carbon isotopic ratio with an uncertainty of $\pm$1.
Table \ref{c1213vel} presents the difference in the theoretical rotational velocity at the ZAMS needed to reproduce observations with these precisions. Thermohaline mixing governs the surface chemical properties of low-mass-RGB stars (M$\lesssim$1.5M$_{\odot}$) after the RGB-bump. Whatever the rotational velocity at the ZAMS, the surface values of $^{12}$C/$^{13}$C of a 1.25 M$_{\odot}$ star during the He-burning phase are almost the same. As a consequence of this, $\Delta$V$_{ZAMS}$ is larger after the RGB than before thermohaline mixing occurs. The efficiency of thermohaline mixing decreases with increasing initial stellar mass \citep{ChaLag10, Lagarde11}. This is the reason why, for intermediate-mass stars, $\Delta$V$_{ZAMS}$ stays almost constant between the beginning of RGB and the He-burning phase. This ideal scenario may be achievable with data from the Kepler satellite which will yield a larger number of targets with precise seismic data However, complementary spectroscopic data \citep[e.g.][]{Carlberg15} with sufficient precision and accuracy will also be necessary.\\

We note however that a discrepancy still exists between the rotation profile deduced from asteroseismic observations  \citep[e.g.][]{Beck12,Mosser12b} and the profiles predicted from models including shellular rotation and related meridional flows and turbulence \citep{Eggenberger12,Marques13}. The rotation rate derived by asteroseimic observations are two orders of magnitude below the rotation rate predicted by theory. This implies the need for a powerful mechanism to extract angular momentum from the core of red-giant stars. More specific informations about the stellar core as the period spacing of g-modes or the core rotation rate could help us to improve stellar models and physical processes occurring in red-giants stars. The surface and core rotation rate as inferred from \textit{Kepler} data will provide additional constraints.\\

With NGC 6633, we presented a first example of a cluster observed by CoRoT including RGB stars, for which chemical properties are also available. It is found that the distances for the cluster members deduced from asteroseismic properties are self consistent, but slightly large compared to Hipparcos distances. Although the stellar masses deduced from seismic properties present significant uncertainties, it is clear that the cluster members are in the mass range where rotation is the most efficient transport processes for chemical elements. Additional information of the rotation profile of these stars is needed to improve our understanding of red-giant stars in this cluster.\\

The space mission \textit{Kepler} and K2 have observed many more open clusters with different turnoff masses, which give us a unique opportunity to follow the evolution of stellar properties through the evolution, and to probe the role of transport processes at different evolutionary phases and different masses.  For many of these stars we will be able to use period spacing and rotational splitting to determine evolutionary state and core rotation rate. To gain the maximum from the data set the asteroseismic properties must be matched by knowledges surface chemical abundances. We have shown in this paper how this complementary data set allows us to provide constraint on the physical processes in stellar interiors. In the future, Gaia-ESO survey, and APOGEE would be extremely helpful. \\

\begin{acknowledgements}
NL  acknowledges financial support from Marie Curie Intra-european fellowship (FP7-PEOPLE-2012-IEF). 
TM achnowledges financial support from Belspo for contract PRODEX GAIA-DPAC.
B.M., C.B., and E.M. acknowledge financial support from the Programme National de Physique Stellaire (CNRS/INSU) and from the ANR program IDEE Interaction Des \'Etoiles et des Exoplan\`etes.
M.R. acknowledges financial support from the FP7 project SPACEINN: Exploitation of Space Data for Innovative Helio- and Asteroseismology.
AM, LG, and EP acknowledge financial support from PRIN INAF-2014.
The research leading to the presented results has received funding from the European Research Council under the European Community's Seventh Framework Programme (FP7/2007-2013) / ERC grant agreement no 338251 (StellarAges).
\end{acknowledgements}

\bibliographystyle{aa}
\bibliography{../../../Bibliographie/Reference}

\end{document}